%% file: main.tex
\documentclass[journal]{IEEEtran}
\IEEEoverridecommandlockouts
\usepackage{cite}    
\usepackage{amsmath,amssymb,amsfonts,upgreek,tabularx}
\usepackage{algorithmic}
\usepackage{textcomp}
\usepackage{xcolor}
\usepackage{cite}
\usepackage{url}
\usepackage{microtype}
\usepackage{amsfonts}
\usepackage{amssymb}
\usepackage{mathrsfs}
\usepackage{euscript}
\usepackage{graphicx}
\usepackage{multirow}
\usepackage{algorithm}
\usepackage{algorithmic}
\usepackage{changepage}
\usepackage{color}
\usepackage{silence}
\usepackage{caption}
\usepackage{subcaption}
\usepackage{todonotes}
\usepackage{scalerel}
\usepackage{tikz}
\usepackage{pgfplots} \usepgfplotslibrary{groupplots} \pgfplotsset{compat=1.18}
\usetikzlibrary{svg.path}
\usepackage{balance}
\usepackage{svg}
\usetikzlibrary{fit, positioning, shapes.geometric, fit, arrows.meta, calc, backgrounds}
\usepackage{pgfplots}
\pgfplotsset{compat=newest}
\definecolor{darkgreen}{rgb}{0, 0.5, 0} %
\definecolor{lightpurple}{rgb}{0.7, 0.4, 1} %
\definecolor{orcidlogocol}{HTML}{A6CE39}
\tikzset{
orcidlogo/.pic={
\fill[orcidlogocol] svg{M256,128c0,70.7-57.3,128-128,128C57.3,256,0,198.7,0,128C0,57.3,57.3,0,128,0C198.7,0,256,57.3,256,128z};
\fill[white] svg{M86.3,186.2H70.9V79.1h15.4v48.4V186.2z}
svg{M108.9,79.1h41.6c39.6,0,57,28.3,57,53.6c0,27.5-21.5,53.6-56.8,53.6h-41.8V79.1z M124.3,172.4h24.5c34.9,0,42.9-26.5,42.9-39.7c0-21.5-13.7-39.7-43.7-39.7h-23.7V172.4z}
svg{M88.7,56.8c0,5.5-4.5,10.1-10.1,10.1c-5.6,0-10.1-4.6-10.1-10.1c0-5.6,4.5-10.1,10.1-10.1C84.2,46.7,88.7,51.3,88.7,56.8z};
}
}
\newcommand\orcidicon[1]{\href{https://orcid.org/#1}{\mbox{\scalerel*{
\begin{tikzpicture}[yscale=-1,transform shape]
\pic{orcidlogo};
\end{tikzpicture}
}{|}}}}
\usepackage[colorlinks=true,linkcolor=blue,citecolor=blue]{hyperref}

\newtheorem{proposition}{Proposition}
\newtheorem{lemma}{\bf{Lemma}}
\newtheorem{corollary}{\bf{Corollary}}
\newtheorem{remark}{Remark}
\newtheorem{theorem}{\bf{Theorem}}
\usepackage[most]{tcolorbox}
\usepackage{booktabs}

\usepackage{xcolor}

\usepackage{bm}
\usetikzlibrary{decorations.pathreplacing}

\usepackage{array}
\AtBeginDocument{\setlength{\abovedisplayskip}{4.5pt plus 2pt minus 2pt}\setlength{\belowdisplayskip}{4.5pt plus 2pt minus 2pt}\setlength{\abovedisplayshortskip}{0pt plus 2pt}\setlength{\belowdisplayshortskip}{4.5pt plus 2pt minus 2pt}}
\input{figs/plotstyle.tex}   %
\newtheorem{assumption}{Assumption}

\newcommand{\E}{\mathbb E}\newcommand{\Var}{\operatorname{Var}}
\newcommand{\cP}{\mathcal P}\newcommand{\cV}{\mathcal V}\newcommand{\cD}{\mathcal D}\newcommand{\cC}{\mathcal C}
\newcommand{\cF}{\mathcal F}\newcommand{\cA}{\mathcal A}\newcommand{\cG}{\mathcal G}
\newcommand{\hh}{\hat{\mathbf h}}\newcommand{\htl}{\tilde{\mathbf h}}\newcommand{\ntl}{\tilde{\mathbf n}}
\newcommand{\mh}{\hat m_h}\newcommand{\mbh}{\bar m_h}\newcommand{\mbd}{\bar m_d}
\newcommand{\ARL}{\mathrm{ARL}_0}

\begin{document}
\title{\huge Label-Free Performance Monitoring of Wireless Foundation Models: Error Estimation, Drift Detection, and Estimator Selection in O-RAN}

\author{
$\text{Mostafa Darabi}^{\orcidicon{0000-0003-4194-3307}}$,\ \IEEEmembership{Member, IEEE}, $\text{Jalal Jalali}^{\orcidicon{0000-0002-3609-6775}}$,\ \IEEEmembership{Member, IEEE}, 
$\text{Melike Erol-Kantarci}^{\orcidicon{0000-0001-6787-8457}}$,\ \IEEEmembership{Fellow, IEEE},\\
and 
$\text{Hina Tabassum}^{\orcidicon{0000-0002-7379-6949}}$,\ \IEEEmembership{Senior Member, IEEE}.
\vspace{-0.0em}
\thanks{
Mostafa Darabi and Jalal Jalali are with the Wireless Communication Research Group, JuliaSpace LLC, Chicago, IL, USA
(e-mail: \{\href{mailto:mosi@juliaspace.com}{\texttt{mosi}},
\href{mailto:josh@juliaspace.com}{\texttt{josh}}\}\texttt{@juliaspace.com}).
Hina Tabassum is with the Department of Electrical Engineering and Computer Science, York University, ON, Canada, M3J 1P3
(e-mail: \href{mailto:hinat@yorku.ca} {\texttt{hinat@yorku.ca}}). 
Melike Erol-Kantarci is with the School of Electrical
Engineering and Computer Science, University of Ottawa, Ottawa, ON K1N 6N5, Canada (e-mail: \href{mailto:melike.erolkantarci@uottawa.ca} {\texttt{melike.erolkantarci@uottawa.ca}}).
\textit{(Corresponding author:
Jalal Jalali.)}
}
}
\markboth{}{}
\maketitle

\begin{abstract}
Wireless physical-layer foundation models (WPFMs) are promising channel estimators, but their accuracy under changing conditions cannot be assessed without ground-truth channel state information (CSI). We develop a label-free monitoring framework for WPFM-based channel estimation in Open Radio Access Networks (O-RAN). Comparing reconstructions of held-out pilots with noisy least-squares observations gives a CSI-free estimate of the channel-estimation mean-squared error (MSE). We characterize its bias and finite-sample concentration, derive conditional pairwise estimator-comparison statistics, and bound the monitored-to-deployed error gap. These results support cumulative sum (CUSUM) drift detection and estimator fallback. The selection guarantees concern held-out error, not deployed estimation error or block error rate (BLER). In 3GPP and ray-traced simulations, the detector meets its false-alarm target in 113 of 135 cases with 100 calibration windows, and in 128 (135 with 200 windows) with tail-aware calibration. Fallback improves normalized MSE (NMSE) by 2.4--4.1\,dB over a frozen WPFM, but MSE-based selection can yield much higher BLER than linear minimum MSE (LMMSE) estimation. A receiver-aware directional statistic lowers this BLER on a switching schedule from 0.33--0.39 to 0.19--0.20 over five runs, still above LMMSE (0.16--0.17). Test-time training increased NMSE after 200 steps, and the slower adaptation loops added little to real-time fallback.
\end{abstract}


\raggedbottom
\section{Introduction}
\IEEEPARstart{W}{ireless} physical-layer foundation models (WPFMs), pretrained on diverse channel datasets, are a promising approach to channel estimation that may generalize beyond the propagation environments seen in training. Unlike task-specific models, WPFMs learn transferable channel representations through self-supervised pretraining, such as masked modeling, and are adapted to tasks and deployments with lightweight heads or low-rank adapters~\cite{alikhani2024lwm,liu2025wifo,cheraghinia2026survey}. In Open Radio Access Networks (O-RAN), they can complement conventional pilot-based channel estimators in the distributed unit (O-DU)~\cite{navidan2026agentic}. Yet WPFMs, like other learned radio functions~\cite{mohajer2026semantic}, remain vulnerable to distribution shifts in delay spread, mobility, blockage, and antenna correlation, and their accuracy cannot be assessed directly during deployment, where ground-truth channel state information (CSI) is unavailable. Key performance measurements (KPMs) such as the block error rate (BLER) indicate estimator degradation only indirectly and with delay, and depend on scheduling, link adaptation, and interference. This motivates label-free monitoring that quantifies the estimation error, detects degradation, and supports estimator-selection and adaptation decisions.

\subsection{Related Work}

LWM~\cite{alikhani2024lwm} and WiFo~\cite{liu2025wifo} use masked modeling to learn transferable channel representations, and surveys~\cite{cheraghinia2026survey,zhang2026multimodal} review WPFM architectures, multi-modal foundation models, adaptation strategies, and hardware considerations. Cheng and Liao~\cite{cheng2026howbig} study pilot-assisted test-time training (TTT) under known scenario changes, and ChannelMAE~\cite{wang2026channelmae} uses masked self-supervision to adapt neural channel estimators online. These works address representation learning and adaptation, not run-time error monitoring and estimator selection.

In a similar approach to ours, Ali, Nasir, and da Costa~\cite{ali2026cognitive} combine pilot-reconstruction scores and neural activation drift to continue, update, or retire a channel estimator, with an LS fallback. Tziouvaras et al.~\cite{tziouvaras2025labelfree} assess concept drift in wireless applications without labels. 
These works show that learned radio functions can be monitored without ground-truth CSI.
However, these studies do not characterize
the bias and finite-sample uncertainty of
pilot-based error monitoring, establish
statistical guarantees for estimator
comparison and fallback, or quantify the
gap between monitored and deployed errors.
These distinctions motivate our framework.

At the network level, agentic O-RAN architectures infer model staleness from unexplained KPM degradation and fall back to conventional radio functions~\cite{navidan2026agentic}. Moreover, dApps support low-latency inference and control in the O-DU~\cite{doro2022dapps,lacava2025dapps,ngrg2024dapps,ngrg2025dapps}. ADORN~\cite{subudhi2026adorn} and RANPilot~\cite{yu2026ranpilot} address drift-aware retraining and adaptation to network reconfigurations, and~\cite{mohajer2026twotimescale} splits O-RAN orchestration between the non-RT RIC and a near-RT xApp. 3GPP also lists model monitoring, updating, and fallback among the AI/ML lifecycle functions~\cite{3gpp38843,lin2023overview,guo2024csi}. These works provide mechanisms for managing AI-based radio functions, but not a statistically characterized, label-free error signal for coordinating estimator selection, drift detection, and adaptation across O-RAN control timescales.

From a statistical perspective, Stein's unbiased risk estimate (SURE)~\cite{stein1981sure}, Noise2Self~\cite{batson2019noise2self}, and hold-out cross-validation~\cite{ward2009cv} estimate prediction risk without clean labels. Harmful distribution shifts have been detected with learned error estimators~\cite{amoukou2024harmful}, prediction risk has been monitored with few labels and finite-sample guarantees~\cite{zhang2026pprm}, and conformal prediction has been used to calibrate uncertainty in wireless systems~\cite{11095417,cohen2022conformal,yoo2025meta,su2026posthoc}. Classical sequential change detection, including the cumulative sum (CUSUM) test~\cite{page1954cusum,lorden1971procedures,moustakides1986optimal,siegmund1985sequential,tartakovsky2014sequential}, detects persistent degradation. We build on these principles.

\subsection{Motivation and Contributions}

Reliable deployment decisions require more
than detecting changes in channel conditions:
they require an observable measure of estimation
error and a characterization of its uncertainty.
We construct such a measure by withholding
selected pilots and correcting their reconstruction
errors for observation noise. While this
hold-out principle is established in
Noise2Self~\cite{batson2019noise2self}
and cross-validation, our contribution is
its statistical characterization and use
for estimator selection, degradation detection,
and adaptation decisions in O-RAN.

\begin{enumerate}

\item \textbf{Statistical characterization of
label-free monitoring:}
We derive the bias, concentration, and
finite-sample accuracy of held-out pilot
error estimates and a conditional
Gaussian comparison of candidate estimators.
We further bound the monitored-to-deployed
error gap, separating measurable terms
from those needing further assumptions
(Lemmas~1--3, Corollary~1, Proposition~1).

\item \textbf{Drift detection and estimator selection:}
Here, drift is a change in the channel or noise statistics (abrupt, gradual, or in SNR) that increases estimation error, detected from that error rather than the inputs. We develop a CUSUM detector with false-alarm and detection-delay guarantees under stated assumptions, a design rule, and a tail-aware calibration for when they fail (Theorem~\ref{thm:detection} and Corollary~\ref{cor:design}). We derive finite-sample selection-error bounds for WPFM/LMMSE fallback, cached-adapter retrieval, and TTT update acceptance (Theorem~\ref{thm:selection} and Corollaries~\ref{cor:buffer}--\ref{cor:ttt}). These bounds concern the realized held-out error, not the deployed error or BLER. We also derive a label-free directional error statistic for receivers on which lower MSE does not mean lower BLER.

\item \textbf{Timescale-aware O-RAN integration:}
We propose a timescale-aware O-RAN architecture coordinating real-time estimator selection, near-real-time degradation detection, and non-real-time adaptation policies.  A cost-aware model gives the minimum shift duration that justifies retrieval, adaptation, or retraining (Proposition~\ref{prop:action}). It decides whether an update is worthwhile, and an acceptance test rejects failed updates, but neither guarantees that adaptation helps. Inference costs are projected analytically onto reference accelerators. Real-time O-DU execution is not validated experimentally.

\item \textbf{Simulation-based validation:}
Using Sionna, we evaluate the framework under 3GPP CDL/TDL and ray-traced channels with abrupt and gradual distribution shifts~\cite{10683207,10556024,hoydis2022sionna}. 
Across five runs, fallback reduces NMSE by 2.4--4.1\,dB relative to a frozen WPFM. Tail-aware calibration raises the detection cells that meet the false-alarm target from 113 to 128 of 135 with the same 100 calibration windows. On a switching schedule, MSE-based selection gives BLER 0.33--0.39 versus 0.16--0.17 for LMMSE, and the directional statistic lowers it to 0.19--0.20, still above LMMSE. TTT increased NMSE after 200 steps, and the full framework changed mean NMSE by at most 0.10\,dB relative to real-time selection. These results do not show that adaptation helps. The five runs share some channel realizations.

\end{enumerate}

Proofs are in the appendices.

\section{System Model and Assumptions}\label{sec:model}

\subsection{Channel Model and Estimators}

We consider the uplink of a single cell, where the gNB has $M$ receive antennas. Each slot contains $L$ OFDM symbols and $K$ subcarriers. We consider one scheduled single-layer UE per time-frequency allocation. Multi-user DMRS code-division multiplexing is discussed in Section~\ref{sec:limits}. The channel over a slot is represented by $\mathbf H\in\mathbb C^{L\times K\times M}$, with $\mathbf h_r\in\mathbb C^M$ denoting the channel at RE $r=(\ell,k)$ of the resource grid $\cG$. Let $\cP\subset\cG$ denote the DMRS REs and $\cD\subseteq\cG\setminus\cP$ the data REs. 
For a unit-modulus pilot $|x_p|=1$, the LS observation is:
\begin{equation}
\htl_p=\mathbf y_px_p^*
=\mathbf h_p+\ntl_p,\qquad
\ntl_p\sim\mathcal{CN}(\mathbf 0,\sigma^2\mathbf I_M).
\end{equation}

A WPFM backbone $f_\theta$, pretrained through masked modeling~\cite{he2022mae}, optionally incorporates a low-rank adapter $\Delta_a$, $a\in\cA$~\cite{hu2022lora}. Given the LS observations on a visible pilot set $\cV\subseteq\cP$, the estimator produces
$\hat{\mathbf H}^{(\cV)}
=f_{\theta,\Delta_a}(\{\htl_p\}_{p\in\cV})$,
with pilot observations outside $\cV$ masked. In deployment, $\cV=\cP$, and a separate reduced-visibility pass serves only monitoring.

The candidate estimators also include LS interpolation and LMMSE interpolation, with the latter using covariance information estimated from visible or previously observed pilots. Let $\cC$ denote the candidate set and $\hat{\mathbf H}^{c,(\cV)}$ the estimate produced by candidate $c\in\cC$ using visible pilots $\cV$.

\subsection{Label-Free Monitoring Framework}

In each monitored slot, we select a subset $\cP_h\subset\cP$ of $n$ pilot REs independently of the observation noise and define $\cV=\cP\setminus\cP_h$ (Fig.~\ref{fig:mask}). Every candidate uses only the visible pilots. Under A1--A2, the withheld-pilot noise is independent of the estimates. Hence, the withheld noisy LS observations serve as an evaluation signal without ground-truth CSI.

We define the noise-corrected monitoring statistic and the paired-estimator comparison as:
\begin{align}
S&=\frac{1}{nM}\sum_{p\in\cP_h}
\big\|\hh^{(\cV)}_p-\htl_p\big\|^2
-\hat\sigma^2,\label{eq:S}\\
S_{AB}&=\frac{1}{nM}\sum_{p\in\cP_h}
\Big(
\big\|\hh^{A,(\cV)}_p-\htl_p\big\|^2
-\big\|\hh^{B,(\cV)}_p-\htl_p\big\|^2
\Big).\label{eq:SAB}
\end{align}
Because the paired statistic scores both candidates on the same withheld pilots, the noise-variance correction cancels.

Let $\mathbf e_p=\hh^{(\cV)}_p-\mathbf h_p$. We distinguish three error quantities. The \emph{realized held-out MSE},
$\mh=\frac{1}{nM}\sum_{p\in\cP_h}\|\mathbf e_p\|^2$,
measures the reduced-visibility estimation error on the withheld REs of the current slot and is unobservable. Its expectation, $\mbh=\E\mh$, is the \emph{expected held-out MSE}. The \emph{deployed data-RE MSE}, $\mbd=\frac{1}{|\cD|M}\sum_{r\in\cD}\E\|\hh^{(\cP)}_r-\mathbf h_r\|^2$, measures the accuracy with all pilots available. Unlike MSE, BLER also depends on the error structure and on receiver processing.

The statistic $S$ estimates the held-out MSE, with bias determined by the noise-variance estimate (Lemma~\ref{lem:unbiased}). It does not directly estimate $\mbd$. Proposition~\ref{prop:gap} characterizes the relationship between monitored and deployed errors.

Two further quantities are observable: the estimator disagreement $\hat\rho_{AB}=\frac{1}{nM}\sum_{p\in\cP_h}\|\hh^{A,(\cV)}_p-\hh^{B,(\cV)}_p\|^2$ and the visibility gap $\hat g_d=\frac{1}{|\cD|M}\sum_{r\in\cD}\|\hh^{(\cP)}_r-\hh^{(\cV)}_r\|^2$, with $\bar g_d=\E\hat g_d$. The error-to-noise ratio is $\eta=\mbh/\sigma^2$.

\emph{Monitoring windows and control loops.}
Window $t$ contains $W$ monitored slots, with window-level statistics $S_t$, $\hat m_{h,t}$, and $S_{AB,t}$. The window independence assumption (A4) is used for the analytical guarantees and is approximated in the simulation through independent UE drops. It is not established for persistent UE trajectories. A regime denotes a stationary distribution of the channel process. Regime changes occur at unknown windows $t_c$. The relative increase in expected held-out MSE is $\kappa=\mbh^{(1)}/\mbh^{(0)}$.

The noise variance is estimated from $k_g$ independent noise-only complex samples per window. They come from REs that carry no scheduled-UE signal and are assumed free of interference, such as guard subcarriers or unused DMRS CDM resources whose noise-only status is ensured. These observations are not used as pilots, enter the estimators only through $\hat\sigma^2$, and are independent of the withheld-pilot noise. In the simulations, $12M$ such samples are generated per slot (Section~\ref{sec:results}). Under the ideal Gaussian model, $\hat\sigma^2\sim\frac{\sigma^2}{2k_g}\chi^2_{2k_g}$, which is unbiased. In practice, interference or leakage may bias it (A3), and unbiasedness of $S$ requires an unbiased estimate.

\begin{figure}[t]\centering\resizebox{\columnwidth}{!}{\input{figs/fig1_arch.tex}}
\caption{Three loops on O-RAN. Only scalar statistics cross the E3/E2 interfaces.}\label{fig:arch}\end{figure}
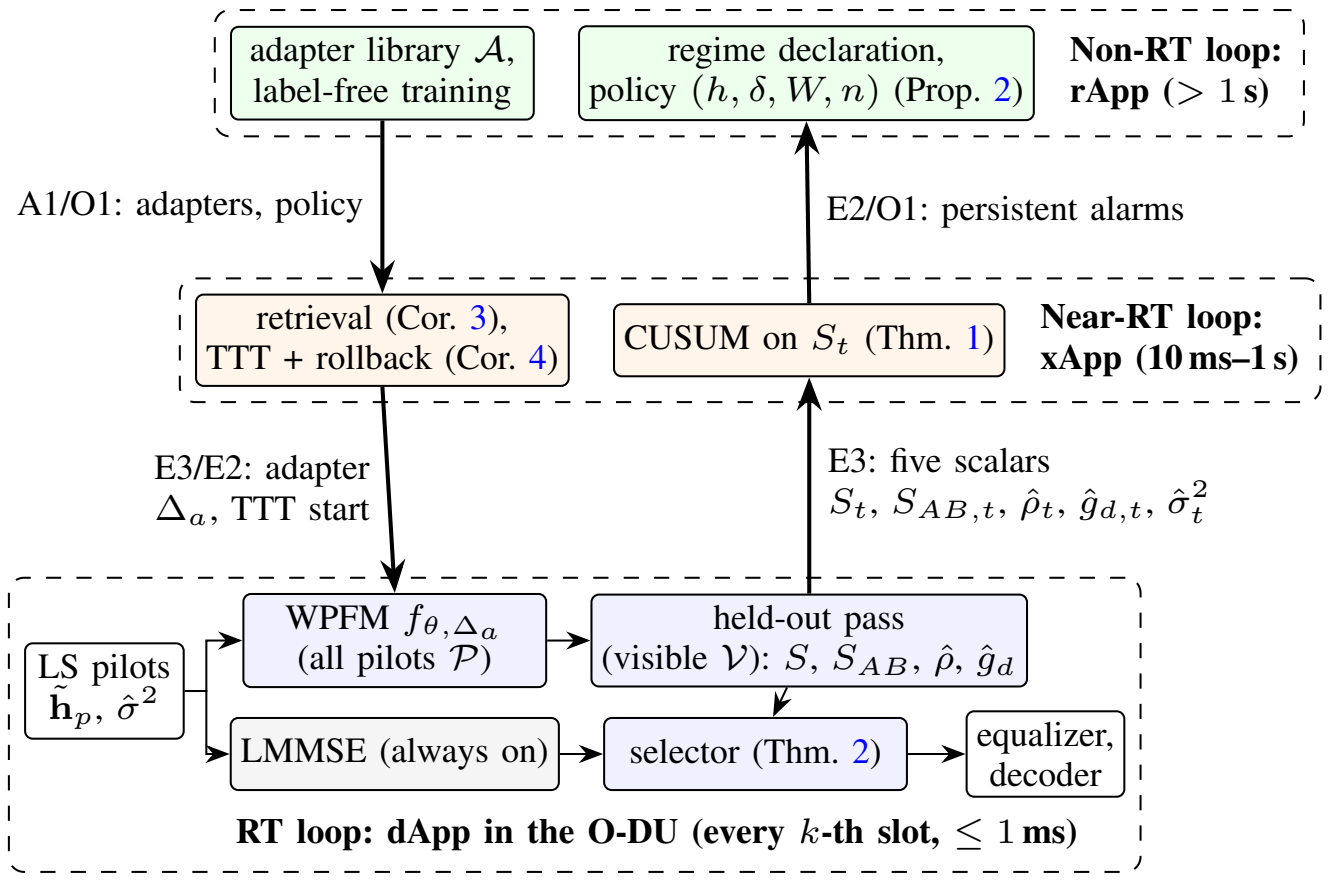

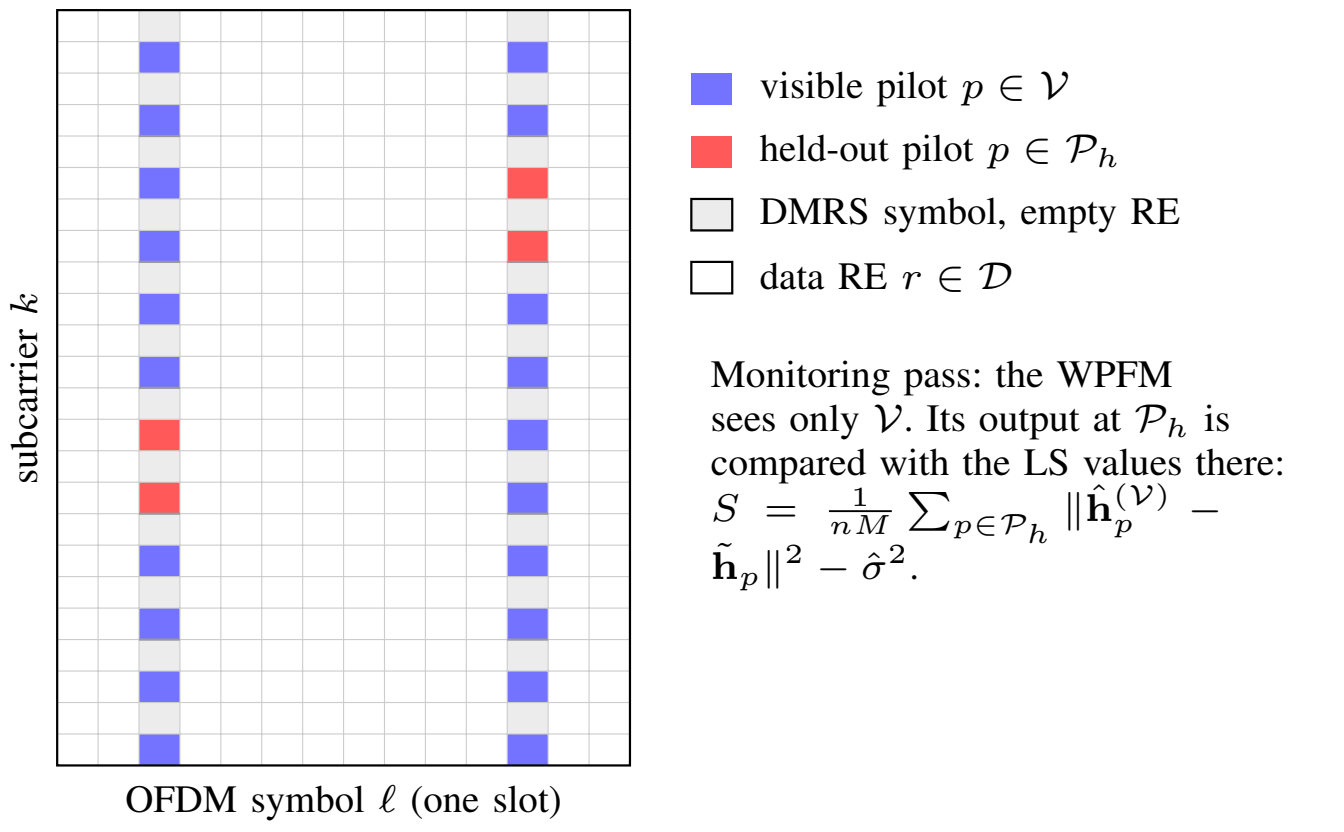
\begin{figure}[t]
\centering
\resizebox{\columnwidth}{!}{\input{figs/fig2_mask.tex}}
\caption{Held-out monitoring on the DMRS grid. All $M$ antenna observations of a held-out token are excluded from the monitoring estimator, while the deployed estimator uses all available pilots. This is the held-out pass of the RT loop in Fig.~\ref{fig:arch}.}
\label{fig:mask}
\end{figure}

The RT loop selects the active candidate $c(t)$, the near-RT loop performs sequential drift detection and responds to alarms, and the non-RT loop manages regimes, adapter training, and the monitoring policy $(h,\delta,W,n)$ (Fig.~\ref{fig:arch}).

\subsection{Statistical Assumptions}\label{sec:assump}

\begin{assumption}[Noise]
Conditioned on the channel process, the pilot noises $\{\ntl_p\}$ are independent and distributed as $\mathcal{CN}(\mathbf 0,\sigma^2\mathbf I_M)$ across REs, antennas, and slots. The noise variance $\sigma^2$ is constant within each monitoring window.
\end{assumption}

\begin{assumption}[Hold-out isolation]
The held-out set $\cP_h$ is selected independently of the observation noise. Every candidate estimate in a window is measurable with respect to:
\[
\begin{aligned}
\cF=\sigma(&\mathbf H,\text{visible-pilot noise},\text{patterns},\\
&\text{earlier held-out noise},\theta,\{\Delta_a\},\xi),
\end{aligned}
\]
where the side information $\xi$ is independent of the current held-out noise. Thus, no candidate uses the current window's withheld LS observations, and LMMSE covariance estimation and TTT updates use only visible or earlier observations.
\end{assumption}

\begin{assumption}[Noise-variance estimation]
The noise-variance estimate $\hat\sigma^2$ is independent of the current held-out noise conditioned on $\cF$. Its bias, defined as
$b_\sigma=\sigma^2-\E\hat\sigma^2$,
is constant within a regime.
\end{assumption}

\begin{assumption}[Window regularity]
Within a stationary regime and for a fixed monitoring configuration and estimator policy, the window statistics $S_t$ are i.i.d., $\hat m_{h,t}\le m_{\max}$, and $\hat m_{h,t}-\mbh$ is sub-Gaussian with proxy $\nu_m^2$.
\end{assumption}

\begin{assumption}[Location transfer]
For a fixed masking pattern, the expected reduced-visibility error $m_r=\frac{1}{M}\E\|\hh^{(\cV)}_r-\mathbf h_r\|^2$ is $L_m$-Lipschitz over $(\cG,d)$, where $d$ denotes RE distance normalized by the channel coherence time and bandwidth.
\end{assumption}

\noindent\emph{Role and scope of the assumptions.}
A1--A3 establish the independence conditions underlying label-free error estimation and paired comparisons (Lemmas~\ref{lem:unbiased}--\ref{lem:paired} and Theorem~\ref{thm:selection}(a)). Unbiasedness additionally requires $\E\hat\sigma^2=\sigma^2$, and conditional unbiasedness requires $\E[\hat\sigma^2\mid\cF]=\sigma^2$. A4 supports the detection analysis (Lemma~\ref{lem:psi}, Theorem~\ref{thm:detection}, and Corollary~\ref{cor:design}) and Theorem~\ref{thm:selection}(b), while A5 supports only the location-transfer components of Proposition~\ref{prop:gap}(ii)--(iii).

A1--A3 are satisfied by construction in the idealized simulations, including the SNR-change scenario in which $\sigma^2$ varies between windows. In practice, colored interference or contaminated noise-only resources may violate them (Section~\ref{sec:limits}). A4 is assumed, not enforced. Because $\hat m_{h,t}$ is unobservable, our implementation can neither clip it at $m_{\max}$ nor impose its tail. The bound $m_{\max}$ enters only the closed-form proxy $\omega_0$ of Lemma~\ref{lem:psi} and Corollary~\ref{cor:design}. Our thresholds use the exact $\psi$ with $\mbh$ and $\nu_m^2$ estimated from calibration windows, which needs the independence and sub-Gaussian parts of A4 but not $m_{\max}$. Both parts can fail: the windows of a persistent UE are dependent, and at high SNR the window statistic is heavier-tailed than the calibrated proxy allows (Section~\ref{sec:res-det}). The detection guarantees hold only where A4 does. Elsewhere, we check the false-alarm rate empirically and calibrate the threshold to the observed tail (Section~\ref{sec:res-det}).

Finally, A5 is a regularity condition. Its Lipschitz constant $L_m$ is estimated with ground-truth channels in simulation and is unavailable during deployment. Consequently, the location-transfer bounds are conditional, not run-time guarantees.

Table~\ref{tab:notation} summarizes the notation. Channel coefficients are normalized so that MSE and NMSE coincide.

\begin{table}[t]
\centering
\caption{Notation}
\label{tab:notation}
\scriptsize
\setlength{\tabcolsep}{2pt}
\renewcommand{\arraystretch}{0.9}
\begin{tabularx}{\columnwidth}{@{}lX@{}}
\toprule
Symbol & Meaning \\
\midrule
$M,K,L$ & Antennas, subcarriers, symbols per slot \\
$\cP,\cP_h,\cV,\cD$ & Pilot, held-out, visible, data RE sets \\
$W,k_g$ & Slots/window, guard samples/window \\
$\sigma^2,\hat\sigma^2,b_\sigma$ & Noise variance, estimate, bias \\
$S,S_t,S_{AB}$ & Held-out, window, paired statistics \\
$\mh,\mbh,\mbd$ & Realized/expected held-out, deployed MSE \\
$\hat\rho_{AB},\hat g_d$ & Disagreement, visibility gap \\
$\eta,\kappa$ & Error-to-noise ratio, MSE inflation \\
$\mu_0,\Delta\mu,v_1$ & Pre-change mean, shift, post-change variance \\
$\nu_0^2,\alpha_0,\omega^\star$ & Sub-exponential parameters, adjustment coefficient \\
$h,\delta$ & CUSUM threshold, design shift \\
$\ARL,D$ & False-alarm run length, detection delay \\
$\Delta,\zeta$ & Selection margin, target error probability \\
$T,\tau_a,C_a,\lambda$ & Dwell time, delay, compute, price \\
$N_\theta,N_{\rm tok}$ & Parameters, tokens per pass \\
\bottomrule
\end{tabularx}
\end{table}

\section{Label-Free MSE Estimation}\label{sec:stat}
We characterize the monitoring signal: its bias (Lemma~\ref{lem:unbiased}), its concentration around the realized error (Lemma~\ref{lem:conc}), the paired form comparing two candidates (Lemma~\ref{lem:paired}), and its relation to the deployed error (Proposition~\ref{prop:gap}).
\begin{lemma}[Unbiasedness]\label{lem:unbiased}
Under A1--A3, $\E[S\mid\cF]=\mh+\sigma^2-\E[\hat\sigma^2\mid\cF]$ and $\E S=\mbh+b_\sigma$. Hence $S$ is unbiased for $\mbh$ if and only if $\E\hat\sigma^2=\sigma^2$, and conditionally unbiased for $\mh$ if and only if $\E[\hat\sigma^2\mid\cF]=\sigma^2$.
\end{lemma}
\noindent The proof (Appendix~\ref{app:stat}) is the J-invariance argument of~\cite{batson2019noise2self}. Conditional unbiasedness is the stronger requirement. In our implementation, $\hat\sigma^2$ is also an input to LMMSE and to the WPFM, hence part of $\xi$ in A2 and $\cF$-measurable. Thus, $\E[S\mid\cF]=\mh+\sigma^2-\hat\sigma^2$, and only unconditional unbiasedness holds. The conditional results below are therefore stated for $S^\circ$, the statistic computed with the true $\sigma^2$. Note that $S=S^\circ-(\hat\sigma^2-\sigma^2)$ exactly. A constant bias of $\hat\sigma^2$ thus shifts the CUSUM baseline, which in-situ calibration absorbs. However, because the design rule derives $\delta$ from $\mu_0$, a bias comparable to $\mbh$ corrupts $\delta$ and $h$ (Section~\ref{sec:results}). A bias that changes after calibration is covered by Remark~\ref{rem:sigma}.

\begin{lemma}[Concentration]\label{lem:conc}
Assume A1--A3 and let $N=nMW$. (a) Given $\cF$, $S^\circ_t-\hat m_{h,t}=G+C$ with $G\sim\mathcal N(0,2\sigma^2\hat m_{h,t}/N)$ and $C=\frac{\sigma^2}{2N}(\chi^2_{2N}-2N)$ uncorrelated, and for every $x>0$, with probability at least $1-4e^{-x}$,
\begin{equation}
|S^\circ_t-\hat m_{h,t}|\le2\sigma\sqrt{\hat m_{h,t}x/N}+\sigma^2\big(\sqrt{2x/N}+x/N\big).
\end{equation}
For the guard estimator add $\sigma^2(\sqrt{2x/k_g}+x/k_g)$ and replace $4e^{-x}$ by $6e^{-x}$. (b) Unconditionally,
\begin{equation}
\Var(S_t)=\frac{\sigma^4+2\sigma^2\mbh}{nMW}+\frac{v_m}{W_{\rm eff}}+\Var(\hat\sigma^2_t),
\end{equation}
where $v_m$ is the per-slot variance of the realized error and $W_{\rm eff}\le W$ accounts for its autocorrelation.
\end{lemma}
\noindent Temporal correlation of the channel reduces only the channel term. The noise term keeps the full count $nMW$ because held-out noise is independent across slots. Correlated estimation errors thus change the variance of the statistic, not its mean, and the variance is measured.

\begin{corollary}[Design rule for $n$]\label{cor:n}
If $\epsilon\eta\le4$ and $nMW\ge\frac{4x}{\epsilon^2}\big(\frac{2}{\sqrt\eta}+\frac{\sqrt2}{\eta}\big)^2$ ($\approx8x/(\epsilon^2\eta^2)$ for $\eta\ll1$), then $|S^\circ_t-\hat m_{h,t}|\le\epsilon\hat m_{h,t}$ with probability at least $1-4e^{-x}$. More pilots stop helping beyond $n_{\rm sat}=(\sigma^4+2\sigma^2\mbh)W_{\rm eff}/(MWv_m)$.
\end{corollary}
\noindent Noise masks model error: a strong estimator at low SNR is hardest to monitor in absolute terms. Comparisons are easier:

\begin{lemma}[Paired comparison]\label{lem:paired}
Under A1--A2, given $\cF$, $S_{AB}\sim\mathcal N\big(\mh^A-\mh^B,\ 2\sigma^2\hat\rho_{AB}/(nM)\big)$ exactly, and $\hat\sigma^2$ cancels. Relative to comparing $S_A$ and $S_B$ on independent hold-out sets of the same size, the variance ratio is at most $(\sqrt{\eta_A}+\sqrt{\eta_B})^2/(1+\eta_A+\eta_B)<2$, and it vanishes as $\eta_A,\eta_B\to0$.
\end{lemma}

\begin{proposition}[Monitoring-to-deployment gap]\label{prop:gap}
Fix a pattern and let $\bar m^{(\cV)}_d$ be the reduced-visibility MSE on data REs. (i) $|\sqrt{\mbd}-\sqrt{\bar m^{(\cV)}_d}|\le\sqrt{\bar g_d}$, with $\bar g_d$ estimable without labels. (ii) Under A5, $|\bar m^{(\cV)}_d-\mbh|\le L_mW_1(U_\cD,U_{\cP_h})$, where $W_1$ is the Wasserstein-1 distance between the uniform distributions on $\cD$ and $\cP_h$ under $d$. (iii) Hence $|\sqrt{\mbd}-\sqrt{\mbh}|\le\epsilon_{\rm gap}:=\sqrt{\bar g_d}+\sqrt{L_mW_1}$. (iv) If $\sqrt{\mbh(r')}-\sqrt{\mbh(r)}>\epsilon_{\rm gap}(r)+\epsilon_{\rm gap}(r')$ for regimes $r,r'$, then $\mbd(r')>\mbd(r)$.
\end{proposition}
\noindent Part (i) is the reverse triangle inequality and (ii) is Kantorovich--Rubinstein duality (Appendix~\ref{app:stat}). Of the gap $\epsilon_{\rm gap}$, only the visibility term $\sqrt{\bar g_d}$ can be evaluated at run time, through its label-free estimate $\hat g_d$. The location term $\sqrt{L_mW_1}$, and with it parts (ii)--(iv), remain theoretical, because $L_m$ is unavailable during deployment. Run-time decisions need the monitored error to preserve the order of the deployed one (iv), not to equal it, and this order is checked empirically (Section~\ref{sec:results}), not guaranteed. Holding out whole DMRS symbols (when two are configured) probes time interpolation and reduces $W_1$ in time. Holding out all antennas of a subcarrier group hides $\mathbf h_p$ entirely and matches patch-masked pretraining~\cite{alikhani2024lwm}.

\section{Drift Detection in the Near-RT Loop}\label{sec:detect}
The xApp runs a CUSUM~\cite{page1954cusum} on window statistics with baseline $\mu_0=\E_0S_t$ calibrated in situ (so it contains $b_\sigma$):
\begin{equation}
G_t=\max(0,G_{t-1}+Z_t),\qquad Z_t=S_t-\mu_0-\tfrac\delta2,
\end{equation}
with $G_0=0$ and alarm time $\tau_A=\inf\{t:G_t>h\}$.
After a change, $\mu_1=\mu_0+\Delta\mu$ with $\Delta\mu=(\kappa-1)\mbh^{(0)}$. Let $\theta_1=\Delta\mu-\delta/2$. (Writing $\mu_1=\kappa\mu_0$ would be wrong whenever $b_\sigma\ne0$.) Two results give the increment's moment generating function. 
It can be shown that $S_t-\mu_0$ is sub-exponential with
\begin{align}
\nu_0^2&=\frac{2\sigma^2(\sigma^2+2m_{\max})}{nMW}+\nu_m^2+\frac{2\sigma^4}{k_g},\\
\alpha_0&=2\sigma^2\max\{1/(nMW),\,1/k_g\}.
\end{align}
This proxy loses about a factor of 2 in the exponent. For design, we therefore use the exact log-MGF.
\begin{lemma}[Adjustment coefficient from the exact noise MGF]\label{lem:psi}
Under A1--A4 with an unbiased guard estimator and $u=\omega\sigma^2/(nMW)<1$, $\ln\E e^{\omega Z_t}\le\psi(\omega)$ with
\begin{multline}
\psi(\omega)=-\omega\big(\mbh+\tfrac\delta2\big)+\frac{\omega\mbh}{1-u}+\frac{\omega^2\nu_m^2}{2(1-u)^2}\\-nMW\ln(1-u)-k_g\ln\big(1+\omega\sigma^2/k_g\big).
\end{multline}
Let $\omega^\star=\sup\{\omega>0:\psi(\omega)\le0\}$. Then $\omega^\star\ge\omega_0:=\min\{\delta/\nu_0^2,1/\alpha_0\}$, and $\omega^\star\to\delta/\Var(S_t)$ as $nMW\to\infty$.
\end{lemma}
\noindent The proof multiplies the exact noncentral chi-square MGF of the held-out residual by the MGF of the guard estimator. The xApp computes the one-dimensional root $\omega^\star$ when it recalibrates.

\begin{theorem}[Detection guarantees]\label{thm:detection}
Under A1--A4: (a) $\ARL=\E_\infty\tau_A\ge e^{\omega^\star h}\ge e^{\omega_0h}$. This part needs only $\E[e^{\omega^\star Z_t}\mid\cF_{t-1}]\le1$, which holds under A4 and, more generally, whenever $\hat m_{h,t}$ is conditionally sub-Gaussian with proxy $\nu_m^2$ around a mean at most $\mbh$ given the past. (b) If $\theta_1>0$ and post-change windows are i.i.d.\ with variance $v_1$, the worst-case delay in Lorden's sense~\cite{lorden1971procedures} satisfies:
\begin{equation}
D\le h/\theta_1+1+v_1/\theta_1^2.
\end{equation}
(c) If $S_t$ is approximately Gaussian with variances $v_0,v_1$, Siegmund's approximation~\cite{siegmund1985sequential} $\mathrm{ARL}(\vartheta)\approx(e^{-2\vartheta\tilde h}+2\vartheta\tilde h-1)/(2\vartheta^2)$ with $\tilde h=h/\sqrt v+1.166$ and standardized drift $\vartheta$ gives design curves.
\end{theorem}
\noindent Part (a) uses the Shiryaev--Roberts-type supermartingale $\Lambda_t=e^{\omega Z_t}(\Lambda_{t-1}+1)$~\cite{pollak1987arl,tartakovsky2014sequential}. Part (b) uses Wald's identity and Lorden's overshoot bound~\cite{lorden1970excess} (Appendix~\ref{app:detect}). Both parts use population values of $\mu_0$, $\mbh$, and $\nu_m^2$. With calibrated estimates, they hold only up to calibration error (Remark~\ref{rem:calib} and~\eqref{eq:tail}).

\begin{corollary}[Design rule]\label{cor:design}
Let $\delta=(\kappa_{\min}-1)\mbh^{(0)}$ and $h=\ln(\ARL^\star)/\omega^\star$. Then $\ARL\ge\ARL^\star$. With $\sigma^2$ known, negligible channel terms, and $\delta\le\sigma^2+2m_{\max}$, the delay satisfies $D\le D^\star$ for every $\kappa\ge\kappa_{\min}$ whenever:
\begin{equation}
nMW\ge\frac{4[(1+2\eta_{\max})\ln\ARL^\star+1+2\kappa_{\min}\eta_0]}{(D^\star-1)(\kappa_{\min}-1)^2\eta_0^2},
\end{equation}
with $\eta_0=\mbh^{(0)}/\sigma^2$ and $\eta_{\max}=m_{\max}/\sigma^2$ (shown with $\omega_0$, and $\omega^\star$ only lowers the requirement). Channel-driven variability adds the floor $(2\nu_m^2\ln\ARL^\star+4\Var(\hat m^{(1)}_{h,t}))/\delta^2$, which more pilots cannot remove.
\end{corollary}
\noindent Since the condition involves $nMW$, the delay in slots, $W(D-1)$, is set by $nM$. Hence, $W$ should be the smallest value that meets the near-RT reporting period and makes windows approximately independent.

\noindent\emph{When A4 fails.} A deployed UE stays for many windows. Since $\nu_m^2$ is calibrated over many drops, a long run of bad drops within the nominal regime is a genuine error excursion, not an operational false alarm. It still counts against $\ARL$, and our one-drop-per-window experiment does not test it. Where model error dominates, $S_t$ can also be heavy-tailed, making its variance an invalid sub-Gaussian proxy. With short calibration, this caused our false-alarm shortfalls (Section~\ref{sec:res-det}).

\begin{remark}[Noise-variance drift]\label{rem:sigma}
If the bias of $\hat\sigma^2$ moves by at most $\bar\varepsilon<\delta/2$ after calibration, (a) holds with $\delta$ replaced by $\delta-2\bar\varepsilon$ and (b) with $\theta_1$ replaced by $\theta_1-\bar\varepsilon$. Section~\ref{sec:results} injects constant $\pm1$ and $\pm3$\,dB errors present during both calibration and test. Drift after calibration is thus not simulated.
\end{remark}
\begin{remark}[Calibration error]\label{rem:calib}
$\mu_0$ is the mean of $C$ nominal windows. On the event $|\mu_0-\E_0S_t|\le\varepsilon_{\rm cal}$, Remark~\ref{rem:sigma} applies with $\bar\varepsilon+\varepsilon_{\rm cal}$ in place of $\bar\varepsilon$. Under A4, Lemma~S3 gives this event probability at least $1-\zeta_{\rm cal}$ for $\varepsilon_{\rm cal}=\max\{\sqrt{2\nu_0^2\ln(2/\zeta_{\rm cal})/C},\,2\alpha_0\ln(2/\zeta_{\rm cal})/C\}$, which needs the unknown $\nu_m^2$ and $m_{\max}$. The implementation instead uses $\varepsilon_{\rm cal}=z\sqrt{\widehat{\Var}(S_t)/C}$ in R3 ($z=2$, sample variance of the same windows): a normal-approximation interval of nominal 95\% coverage, not a finite-sample guarantee, which the calibration error exceeded in one run (Section~\ref{sec:res-det}).
\end{remark}
\begin{remark}[Gradual drift, unknown $\kappa$, SNR]\label{rem:ramp}
For a linear ramp of length $R$ windows, the mean path of $G_t$ crosses $h$ after about $R\delta/(2\Delta\mu)+\sqrt{2Rh/\Delta\mu}$ windows (fluid approximation).
With $J$ parallel charts and $h_j=\ln(J\ARL^\star)/\omega_j$, the earliest alarm still satisfies $\ARL\ge\ARL^\star$.
This costs $\ln J$ per threshold and is a simple alternative to a generalized-likelihood-ratio CUSUM over $\delta$~\cite{tartakovsky2014sequential}. As $\mbh$ depends on SNR, $\mu_0$, $\delta$, and $\nu_0^2$ are indexed by an SNR bin from $\hat\sigma^2$. Part (a) then holds with the smallest $\omega$ over bins.
\end{remark}

\section{Real-Time Fallback and Adapter Retrieval}\label{sec:select}
In the RT loop, a dApp in the O-DU picks the estimator for each window ($\cC=\{\text{WPFM},\text{LMMSE}\}$). After an alarm, the xApp retrieves an adapter from $W$ buffered slots ($\cC=\cA\cup\{\text{LMMSE}\}$). Both use $\hat c=\arg\min_c\sum S_c$. Let $c^\star=\arg\min_c\mh^c$, the candidate of lowest realized held-out error.
\begin{theorem}[Selection error]\label{thm:selection}
(a) Under A1--A2, given $\cF$, with $N=nMW$ and $\hat\rho_{\max}=\max_{c,c'}\hat\rho_{cc'}$,
\begin{multline}
\Pr(\mh^{\hat c}>\mh^{c^\star}+\Delta\mid\cF)\le\frac12\sum_{c:\hat\Delta_c>\Delta}e^{-N\hat\Delta_c^2/(4\sigma^2\hat\rho_{cc^\star})}\\\le\frac{|\cC|-1}2e^{-N\Delta^2/(4\sigma^2\hat\rho_{\max})},
\end{multline}
where $\hat\Delta_c=\mh^c-\mh^{c^\star}$. (b) Under A1, A2, A4, if the window-averaged realized difference is sub-Gaussian with proxy $\nu_\Delta^2$ and $\hat\rho\le\rho_{\max}$, then for expected MSEs the error probability is at most $(|\cC|-1)[\frac12e^{-N\Delta^2/(16\sigma^2\rho_{\max})}+e^{-\Delta^2/(8\nu_\Delta^2)}]$.
\end{theorem}
\begin{corollary}[Buffer size]\label{cor:buffer}
A candidate within $\Delta$ of the best is selected with probability at least $1-\zeta$ if $nMW\ge\frac{4\sigma^2\hat\rho_{\max}}{\Delta^2}\ln\frac{|\cC|-1}{2\zeta}$. If the candidates are fixed over the buffer, it may be stopped adaptively by a rule using only $\hat\rho$, not $S_c$.
\end{corollary}
\begin{corollary}[TTT acceptance and rollback]\label{cor:ttt}
Train adapter $a'$ from $a$ on earlier windows and accept it only if $S_{a'a}\le-\gamma$ on a fresh hold-out set. Then $\Pr(\text{accept}\mid\mh^{a'}\ge\mh^a)$ and $\Pr(\text{reject}\mid\mh^{a'}\le\mh^a-2\gamma)$ are both at most $\frac12e^{-nMW\gamma^2/(4\sigma^2\hat\rho_{a'a})}$. On rejection, $a$ is kept.
\end{corollary}
\noindent Theorem~\ref{thm:selection} bounds the probability of selecting a candidate whose realized held-out error exceeds the smallest by more than $\Delta$. It does not guarantee that the candidate with the lowest monitored error also has the lowest deployed error on the data REs or in the next window. This requires Proposition~\ref{prop:gap}(iv), whose gap terms cannot be evaluated at run time, and a persistent regime. It is supported empirically (Section~\ref{sec:results}) but not guaranteed. Likewise, since LMMSE is always a candidate, the held-out error of the selected candidate exceeds LMMSE's by more than $\Delta$ at most with the probability of Theorem~\ref{thm:selection}(a). This limits the MSE downside of a failed WPFM only as far as held-out and deployed errors agree, and it does not extend to BLER (Section~\ref{sec:results}). TTT on a held-out loss could overfit the pilot noise of its own training windows. For this reason, Corollary~\ref{cor:ttt} validates on a fresh set. The requirement grows only logarithmically in the library size, and side information (cell, band, time of day) prunes $\cC$.

\section{Adaptation Decisions and Design Rules}\label{sec:action}
The fallback of Section~\ref{sec:select} needs no model update. Retrieval, TTT, and retraining may lower the error further. Let $\ell(m)$ be any nondecreasing per-window loss: $\ell(m)=m$, or BLER at the effective SINR $\gamma_{\rm eff}(m)=M(1-m)\mathrm{SNR}/(1+m\,\mathrm{SNR})$ that treats estimation error as noise~\cite{hassibi2003training} (exact for LMMSE, approximate for neural estimators). Each action $a\in\{\text{keep},\text{retrieve},\text{TTT}(k),\text{retrain}\}$ has a delay $\tau_a$ (windows) with interim MSE $m^{\rm pre}_a$, steady MSE $m_a$, and compute $C_a$ (B1). TTT follows $m(k)=m_\infty+(m_0-m_\infty)\varrho^k$. The dwell time $T$ is independent of the action, and $T\ge\max_a\tau_a$ (B2).
\begin{proposition}[Action selection]\label{prop:action}
Under B1--B2, $\mathcal L_a(T)=\tau_a\ell(m^{\rm pre}_a)+(T-\tau_a)\ell(m_a)+\lambda C_a$. (i) The optimal action is the lower envelope of the lines, with steady loss nonincreasing in $T$. (ii) Actions with $\ell(m_a)>\ell(m_b)$ break even at:
\begin{equation}
T_{ab}=\frac{\lambda(C_b-C_a)+\tau_b(\ell^{\rm pre}_b-\ell_b)-\tau_a(\ell^{\rm pre}_a-\ell_a)}{\ell_a-\ell_b}.
\end{equation}
(iii) With $\ell(m)=m$ and $\tau_k=kt_{\rm s}$, $\mathcal L_{\rm TTT}(k)$ is strictly convex for $kt_{\rm s}<T$, and for $k^\star t_{\rm s}\ll T$, $k^\star\approx\frac{1}{\ln(1/\varrho)}\ln\frac{T(m_0-m_\infty)\ln(1/\varrho)}{\lambda c_{\rm TTT}+t_{\rm s}(m_0-m_\infty)}$ ($k^\star=0$ if the argument is below 1). (iv) For random $T$, (i)--(iii) hold with $\E T$.
\end{proposition}
\noindent The resulting decision rules characterize when adaptation is cost-effective under the assumed performance model. They do not guarantee that a learned update improves estimation accuracy, and Corollary~\ref{cor:ttt} rejects updates that fail. Running LMMSE through the RT selector during an action's delay lowers $m^{\rm pre}$ and every break-even time. The design rules follow:
\textbf{R1} Choose $nMW$ from Corollary~\ref{cor:n} at the smallest $\eta$ of interest, not beyond $n_{\rm sat}$, and cap $n/|\cP|$ so that $\sqrt{\hat g_d}\le\frac12\sqrt{S}$.
\textbf{R2} Use the smallest $W$ that meets the near-RT period and decorrelates windows.
\textbf{R3} Set $\delta=(\kappa_{\min}-1)\mbh^{(0)}-2\bar\varepsilon$ and $h=\ln\ARL^\star/\omega^\star$ per SNR bin, with $J$ charts when $\kappa$ is unknown.
\textbf{R4} Select WPFM or LMMSE every window. After an alarm, score cached adapters (Corollary~\ref{cor:buffer}). Start TTT only if $\E T>T_{\rm retrieve,TTT}$ and $k^\star\ge1$, accepting it only via Corollary~\ref{cor:ttt}. Retrain when $\E T>T_{\rm TTT,retrain}$ or alarms persist.

\section{O-RAN Instantiation and Compute}\label{sec:oran}
Fig.~\ref{fig:arch} and Table~\ref{tab:loops} map the loops onto O-RAN. The LS pilot grid never leaves the O-DU. Only scalars cross E3/E2. The RT loop relies on dApps. The E3 interface is described by nGRG research reports~\cite{ngrg2024dapps,ngrg2025dapps} and by an OpenAirInterface implementation with average control latency below 450\,$\mu$s~\cite{lacava2025dapps}. E3 is not standardized. Without a dApp, the selector can run inside the PHY pipeline while the detector stays in an xApp.

\begin{table}[t]\centering\caption{Loops, components, and interfaces}\label{tab:loops}\scriptsize
\begin{tabular}{@{}>{\raggedright}p{0.9cm}>{\raggedright}p{1.6cm}>{\raggedright}p{3.3cm}>{\raggedright\arraybackslash}p{1.45cm}@{}}\toprule
Loop & Component, period & Runs & Interface\\\midrule
RT & dApp in O-DU, every $k$-th slot & WPFM, held-out pass, $S$, $S_{AB}$, $\hat\rho$, $\hat g_d$, selector (Thm.~\ref{thm:selection}) & E3 (not standardized)\\
Near-RT & xApp, \mbox{10\,ms--1\,s} & CUSUM (Thm.~\ref{thm:detection}), retrieval (Cor.~\ref{cor:buffer}), TTT with rollback (Cor.~\ref{cor:ttt}) & E2, E3\\
Non-RT & rApp/SMO, $>1$\,s & regime declaration, label-free adapter training, library, policy (Prop.~\ref{prop:action}, Cor.~\ref{cor:design}) & A1, O1\\\bottomrule
\end{tabular}\end{table}

\emph{Compute.} One token costs $2N_\theta/T_{\rm ops}$ (compute bound)~\cite[Sec.~III-D]{cheraghinia2026survey}. 
A Transformer pass over $N_{\rm tok}$ tokens costs: 
\begin{equation} 
t_{\rm pass}=
\max\Big(\frac{2N_\theta N_{\rm tok}+4N_{\rm layer}N_{\rm tok}^2d}{T_{\rm ops}},\ \frac{N_\theta b/8}{B_{\rm mem}}\Big), \end{equation} 
with the load term dropped when the $b$-bit weights fit in the usable 10--30\% of on-chip memory~\cite{cheraghinia2026survey}. Monitoring every $k$-th slot adds $t_{\rm pass}/k$ per slot. Per window the dApp sends five scalars ($S_t$, $S_{AB,t}$, $\hat\rho_t$, $\hat g_{d,t}$, $\hat\sigma^2_t$), 20 bytes in single precision. Table~\ref{tab:compute} applies this to the survey's two reference platforms. These are analytical projections, not measurements. 
They assume peak throughput and exclude data movement, scheduling overheads, and contention with other PHY functions. They are therefore lower bounds, and the GPU column must not be read as free compute. On the FPGA, the two passes per slot (one per DMRS symbol) of an LWM-1.1-class model take 43\% of the 500\,$\mu$s budget. Deploying and monitoring every slot uses 86\% of that budget even at peak throughput. Monitoring every fourth slot lowers this value to 54\%. At a similar parameter count, the in-house base model is about 3 times cheaper because its encoder sees 24 coarse pilot tokens (4 subcarriers $\times$ $M$ antennas each) instead of $4\times4$ patches of the antenna--subcarrier grid. 
Our only measurements are the CPU latencies of Table~\ref{tab:timing}, which rank the operations by order of magnitude but are not O-DU latencies. Real-time O-DU execution therefore remains unvalidated.

\begin{table}[t]\centering\caption{Per-slot compute (analytical projection, INT8, peak throughput, no overheads)}\label{tab:compute}\scriptsize\setlength{\tabcolsep}{2.5pt}
\begin{tabular}{@{}lrrrrr@{}}\toprule
& & \multicolumn{3}{c}{FPGA: 5 TOPS, 500\,$\mu$s} & GPU\\\cmidrule(lr){3-5}
Model & $N_\theta$ & mon.\ pass & dep.+mon., $k{=}1$ & $k{=}4$ & $k{=}1$\\\midrule
MAE small (ours) & 0.75\,M & 24.3\,$\mu$s & 49.8 (10\%) & 31.6 (6\%) & 0.40\,$\mu$s\\
MAE base (ours) & 2.61\,M & 65.1\,$\mu$s & 134.9 (27\%) & 86.1 (17\%) & 1.08\,$\mu$s\\
LWM 1.0 class & 0.6\,M & 57.4\,$\mu$s & 114.8 (23\%) & 71.8 (14\%) & 0.92\,$\mu$s\\
LWM 1.1 class & 2.5\,M & 214.6\,$\mu$s & 429.3 (86\%) & 268.3 (54\%) & 3.44\,$\mu$s\\\bottomrule
\multicolumn{6}{@{}p{\columnwidth}@{}}{\scriptsize Grid of Section~\ref{sec:results} ($M{=}32$, $K{=}48$, 24 pilot tokens, 6 held-out tokens). Platforms from~\cite[Fig.~4]{cheraghinia2026survey}: FPGA (32\,MB, 100\,GB/s) and GPU (624 TOPS). In-house rows: monitoring pass over 18 visible tokens (deployed pass 25.5 and 69.8\,$\mu$s). LWM rows~\cite{alikhani2024lwm,wilab2025lwm11}: 12 layers, $d=64$ (1.0) and 128 (1.1), 4$\times$4 patches, one pass per DMRS symbol, both counted. Monitoring pass equal to the deployed pass (12 layers for LWM 1.0 is assumed).}
\end{tabular}\end{table}

\begin{table}[t]\centering\caption{Measured operation latencies on a four-core CPU virtual machine (PyTorch 2.14.1, mean over 50--10\,000 repetitions)}\label{tab:timing}\scriptsize
\begin{tabular}{@{}llr@{}}\toprule
Operation & Budget & Measured [$\mu$s] \\\midrule
WPFM deployed pass (1 slot) & RT, 500\,$\mu$s & 5180\\
WPFM monitoring pass (1 slot) & RT & 5636\\
LMMSE (cached filter, 1 slot) & RT & 188\\
CUSUM update & near-RT, 10\,ms & 0.13\\
Adapter swap & near-RT & 140\\
TTT step (batch 1) & near-RT & 18\,929\\\bottomrule
\end{tabular}\end{table}

\section{Numerical Results and Discussion}
\label{sec:results}

We evaluate whether label-free monitoring can reliably assess the estimation error of the WPFM and improve deployment decisions as conditions change. The experiments cover estimator comparison, detection of harmful degradation, validation of the detection bounds, end-to-end selection, receiver performance, and the benefits and limits of label-free adaptation.

\subsection{Experimental Setup}
\label{sec:res-setup}

\subsubsection{\bf System and Propagation Models}
We use Sionna 2.2.0 (PyTorch backend)~\cite{hoydis2022sionna} to simulate a
3.5\,GHz uplink with 30\,kHz subcarrier spacing. The grid has $K=48$
subcarriers (four PRBs) and $L=14$ OFDM symbols per slot. For CDL channels,
$M=32$ receive antennas are arranged in a $4\times8$ planar array. Type-1 DMRS
occupies symbols 2 and 11 with comb~2. Each slot includes $12M$ guard samples,
yielding $k_g=12MW$ samples per monitoring window. Propagation follows the 3GPP
TR~38.901 CDL and TDL models, which cover variations in channel profile, delay
spread, mobility, and antenna correlation. Unless otherwise specified, each
window contains $W=20$ slots from a new UE drop. Successive windows therefore
represent different UE drops rather than a persistent UE trajectory, except in
the ray-traced scenario.

\subsubsection{\bf WPFM Architecture and Training}
The WPFM is a masked autoencoder with
0.75 million parameters, embedding
dimension $d=96$, four encoder layers,
and two decoder layers. Each token covers four subcarriers across $M$ antennas. The model is pretrained
without labels on noisy CDL-A/B/D and
TDL-A/C channels, using
75\% masking and SNRs from $-5$ to
30\,dB. It is then fine-tuned for
channel estimation on CDL-A/B with
random hold-out augmentation for reduced pilot visibility.

Rank-4 LoRA adapters are applied to attention and MLP projections. The library
includes the base adapter and five adapters, one per regime:
CDL-C (300\,ns), CDL-A (60\,km/h), correlated TDL-A, CDL-B (200\,ns), and CDL-E.
These adapters are trained without labels on logged LS pilots. Pretraining uses 6000 steps, fine-tuning uses 4000 steps with a batch size of 64, and each adapter uses 500 steps. The statistical results in Sections~\ref{sec:stat} to~\ref{sec:select}
apply to any fixed estimator that satisfies A1 to A5, independently of the WPFM architecture.

\subsubsection{\bf Monitoring and Decision Configuration}
We withhold six of the 24 pilot tokens per slot. These are $n=12$
pilot REs and $nM=384$ observations across antennas. Under A1 and A2, the noise
on the withheld pilots is independent of the estimates constructed from the
visible pilots, enabling label-free error monitoring. The near-RT detector
of the end-to-end runs uses $\kappa_{\min}=1.5$, $\ARL^\star=1000$ windows,
$C=50$ calibration windows, and $z=2$. The detection experiments of
Section~\ref{sec:res-det} use $\ARL^\star\in\{50,100,200\}$ and $C=100$, with
$C=200$ in the tail-aware extension and in the test of gradual transitions. The
real-time (RT) selector compares candidates on the same withheld pilots and
accumulates their statistics over window $t$. It then deploys the selected
estimator on all available pilots in window $t+1$. Decisions are evaluated on a new UE drop, not on the observations used for selection.

\subsubsection{\bf Baselines and Framework Configurations}

We compare against conventional, learned, and oracle baselines. Conventional estimators include least squares (LS), LMMSE with the
channel covariance over time and frequency tracked from previous pilots, and genie-aided LMMSE with the true regime statistics. Learned baselines include the frozen WPFM, always-on test-time training (TTT) using the
held-out loss, and periodic label-free retraining every 200 windows. A further baseline retrains when the exponentially weighted moving average (EWMA) of the BLER exceeds 0.15. We also consider supervised estimators, one per regime, each trained from scratch for 800 steps with true channels as labels. They have approximately 7\% more parameters than an equal share of the comparison model budget and are selected by the same paired rule. Access to labels favors them, although shorter training limits the comparison.

A regime oracle that cannot be deployed selects the library adapter or LMMSE
with the lowest true NMSE on a probe of each phase. It provides a reference for
estimator selection.

The framework is evaluated in four configurations: RT selection alone, RT with adapter retrieval, RT with TTT, and the full framework with the non-RT loop of Section~\ref{sec:action}. 
These isolate the contributions of selection, retrieval, and adaptation.

\subsubsection{\bf Evaluation Scenarios}

We consider four temporal scenarios. Unless otherwise
specified, the nominal SNR is 0.5\,dB, where LMMSE
has 7.5\% BLER on CDL-A at MCS~20.

{\textbf{1}) Abrupt regime changes:}
The channel switches every 100 windows through
CDL-A $\rightarrow$ CDL-C (300\,ns)
$\rightarrow$ CDL-A (60\,km/h)
$\rightarrow$ correlated TDL-A
$\rightarrow$ CDL-B (200\,ns)
$\rightarrow$ CDL-A.

{\textbf{2}) SNR changes:}
The channel remains CDL-A and SNR
changes from 10 to 0 to 20\,dB, with
100 windows per condition.

{\textbf{3}) Gradual propagation changes:}
A linear power blend over 200 windows transitions from CDL-A to CDL-C, between two stationary phases of 100 windows each.

{\textbf{4}) Ray-traced blockage:}
A CDL-A phase of 100 windows is followed by 200 windows in a Sionna ray-traced
street canyon, with a moving blocker along a UE trajectory. The regime
boundaries are unknown to the monitoring framework.

\begin{figure*}[t]\centering\input{figs/fig_v02_maps.tex}
\caption{Condition map of the reference run: ten regimes (rows, separated by lines into those used in fine-tuning, those seen only in pretraining, and unseen ones) by 13 SNRs (columns, with those left of the dashed line below the training SNR range). (a) Ratio of the NMSE of the WPFM to that of LMMSE on data REs (blue: WPFM better, dots: the two differ by less than 1\,dB). (b) Its label-free estimate from the summed held-out statistics (crosses: sign differs from (a)). (c), (d) NMSE regret of the proposed selector and of an SNR rule tuned with labels on CDL-A/B, relative to the better estimator of each cell. (e) MSE inflation $\kappa$ of the WPFM relative to CDL-A at the same SNR (dots: $\kappa\ge\kappa_{\min}=1.5$). (f), (g), (h) Alarm rate per window of three monitors (dots as in (e)) at a 5\% false alarm rate on CDL-A at the same SNR.}\label{fig:maps}\end{figure*}
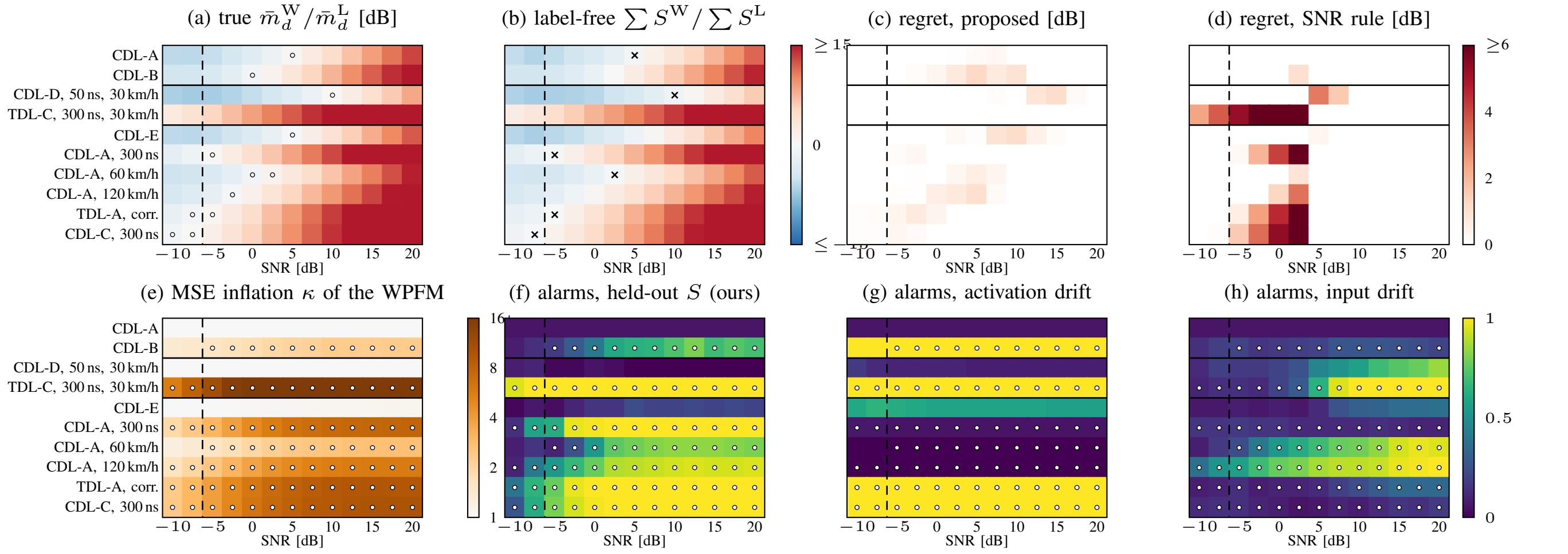

\emph{1) Condition map:}
The frozen WPFM and tracked LMMSE are compared across ten propagation regimes and
13 SNRs ($-10$ to 20\,dB). These include two fine-tuning regimes (CDL-A/B), two
seen only during pretraining, and six unseen regimes. Each condition uses 60
windows (150 for CDL-A). Channel realizations are shared and the noise is
normalized across SNRs within each regime, which enables paired comparisons
(Fig.~\ref{fig:maps}).

\emph{2) Link map:}
We evaluate 13 MCS values (3 to 27) on CDL-A, CDL-E, and CDL-C (300\,ns) over
SNRs from $-20$ to 5\,dB. Each point uses ten windows. Decoding with the WPFM, LMMSE, or true channel uses the same bits and noise.

\emph{3) Directional evaluation:}
The four scenarios are repeated in five runs with only the two RT candidates and ten transport blocks per window. These runs use channel realizations different from the main evaluation and are reported separately.

BLER is computed with the Sionna 5G NR PUSCH chain (MCS Table~1 of 3GPP TS~38.214), treating estimates as exact.

\subsubsection{\bf Reproducibility}

We conduct five runs with different initialization and training seeds. The
reference run (seed 20261026) uses a cloud virtual machine with four cores,
while seeds 1 to 4 use an Apple M3 Max with identical code and package
versions. Because some channel realizations are shared across seeds 1 to 4 (Section~S-VII), the runs are not independent channel experiments and underestimate channel variability. The condition and link maps, the hold-out
augmentation ablation, the operating point sweep, and the diagnostics of error
direction and tail delay use only the reference run. The figures do as well, unless stated otherwise. Experiments repeated across five runs report the range or mean.

\subsection{Label-Free Estimator Comparison and Selection}
\label{sec:res-map}

\subsubsection{\bf Estimator selection}
Fig.~\ref{fig:maps}(a) shows that the relative performance of the WPFM and the
tracked LMMSE depends strongly on SNR and propagation conditions. The WPFM
benefits from its learned channel prior at low SNR but reaches an error floor
($-18.6$ to $-18.3$\,dB on CDL-A at 20\,dB). The LMMSE error decays approximately as $1/\mathrm{SNR}$. The WPFM outperforms LMMSE in 40 of 130 combinations of regime and SNR, with crossovers in nine of ten regimes. For example, the WPFM is preferable up to
2.5\,dB on CDL-A and up to 7.5\,dB on CDL-D. Larger delay spread and higher
mobility shift the crossover to lower SNRs. Familiarity with the training set
is not a reliable indicator of performance. The WPFM is 13.1\,dB worse on CDL-A
at 20\,dB but up to 6.1\,dB better on the unseen CDL-E. These results favor
selecting the estimator by measured performance rather than by regime.

\subsubsection{\bf Accuracy of label-free comparison}
Fig.~\ref{fig:maps}(b) shows that the held-out statistic identifies the estimator with the lower NMSE in 124 of 130 conditions. The six incorrect
decisions occur near crossovers, where the true NMSE difference is below
0.9\,dB. This is consistent with the dependence on the gap in
Theorem~\ref{thm:selection}. Across 42 additional conditions, the measured mean
and variance agree with Lemmas~\ref{lem:unbiased} and \ref{lem:conc}(b), with
median deviations within 1.5\%.
In the reference run, the median normalized visibility gap $\hat g_d/\mbd$ is 0.030 with and 0.032 without hold-out augmentation.
These observations support the accuracy of held-out monitoring under the tested conditions but establish no universal guarantee relating the monitored error to the deployed error.

The hold-out is essential to avoid evaluating an estimator on the noisy pilots
used to construct it. The in-sample residual favors LMMSE in 129 of 130 conditions, including 39 of the 40 in which the WPFM wins. It causes a selection loss of up to 6.8\,dB. Independent held-out observations avoid this bias.

\subsubsection{\bf Selection performance}
Figs.~\ref{fig:maps}(c) and (d) show the NMSE regret of the selector, which
chooses the estimator for window $t+1$ from the statistics of window $t$. The
mean regret is 0.12\,dB, with a maximum below 1.3\,dB. An SNR threshold calibrated with labels on CDL-A/B incurs up to 10.3\,dB regret in other regimes. A fallback based on activation drift reaches 21.3\,dB because it
confuses unfamiliar channels with harmful estimation errors. Accumulating the
monitoring statistic over five windows reduces the maximum regret to 0.87\,dB at
the cost of a slower response to regime changes. Held-out error monitoring thus supports estimator selection with low regret in deployed NMSE, without true CSI or knowledge of the regime. BLER is examined in Section~\ref{sec:res-link}.

\subsection{Estimation Degradation Versus Distribution Shift}
\label{sec:res-harm}

\subsubsection{\bf Distribution shift versus estimation degradation}
Fig.~\ref{fig:maps}(e) shows the MSE inflation factor $\kappa$ of the WPFM
relative to CDL-A at the same SNR. CDL-E is unseen during training but exhibits
negligible degradation ($\kappa=0.99$ to $1.09$). CDL-D, seen only during pretraining, has $\kappa<1$. Conversely, $\kappa$ exceeds the
detection threshold of 1.5 on the fine-tuning regime CDL-B at SNRs above
$-5$\,dB and on several unseen regimes. Degradation also varies with SNR. On
CDL-A at 60\,km/h, $\kappa$ increases from 1.2 to 2.7 as the SNR rises from
$-10$ to 20\,dB. Thus, neither familiarity with the training set nor
distribution shift reliably indicates estimation degradation.

\subsubsection{\bf Detection performance}
Figs.~\ref{fig:maps}(f) to (h) compare the held-out monitor with activation and input drift detectors, all calibrated to a 5\% false
alarm rate on CDL-A. Activation drift flags 59\% of unseen CDL-E windows despite
negligible MSE inflation, but misses the degradation caused by mobility on CDL-A. Input drift flags 43\% of benign CDL-D windows but only 13\% of
severely degraded CDL-A (300\,ns) windows.

Across the evaluated conditions, the held-out monitor alarms in 82\% of harmful
windows ($\kappa\geq1.5$) and 9\% of benign windows ($\kappa\leq1.2$). The rates are 58\%/34\% for activation drift and 38\%/31\% for input
drift. Averaging over five windows improves the held-out monitor to 96\%/8\%,
whereas the benign alarm rates of the drift detectors increase to 60\% and
53\%, respectively (Fig.~S3). The 5\% calibration target applies to CDL-A, and
these aggregate rates cover multiple regimes.

These results show that monitoring estimation error, rather than distribution
novelty, provides a more reliable signal for detection and adaptation decisions
based on performance.

\begin{figure}[t]\centering\input{figs/fig_v02_tail.tex}
\caption{Empirical $\ARL/\ARL^\star$ of the 135 detection cells (five runs) with the plug-in threshold of Corollary~\ref{cor:design} and the tail-aware threshold~\eqref{eq:tail}, both from $C=100$ calibration windows. Values at 2.5, 5, and 10 are lower bounds (no false alarms in 500 windows). Bottom: cells that meet the target, plug-in$\,\rightarrow\,$tail-aware.}\label{fig:tail}\end{figure}
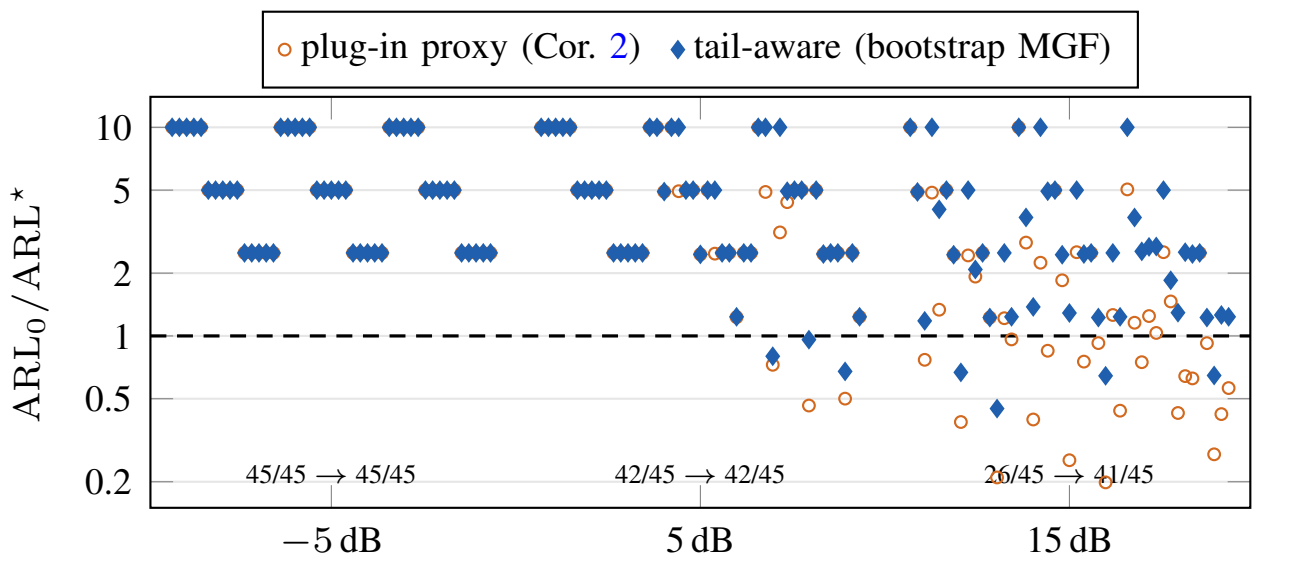

\subsection{Validation of Detection Bounds and Tail-Aware Calibration}
\label{sec:res-det}

We evaluate the CUSUM guarantees using 100 CDL-A calibration windows (the
end-to-end detector uses $C=50$). For false alarms, the test streams have 500 windows with 50 staggered starts. Detection delays are
evaluated after switching to five CDL-C power blends ($\kappa=1.80$ to $11.5$),
using streams of 100 windows.
Three SNRs ($-5$, 5, and 15\,dB), three hold-out sizes (2, 4, and 8 tokens), and three targets ($\ARL^\star\in\{50,100,200\}$) yield 27 configurations per run and 135 total.

\subsubsection{\bf Detection Delay}
Measured delays satisfy Theorem~\ref{thm:detection}(b) whenever all starts
trigger within 100 windows. The median ratio of delay to bound is 0.74 to 0.78 across runs. In the other 22 cases, all at $-5$\,dB with $\kappa\leq2.64$,
the bound exceeds the observation horizon, preventing validation. Gradual transitions lasting 10 to 200 windows are detected
(with $C=200$ and $\ARL^\star=200$). Delays are 0.70 to 0.94 times those predicted by Remark~\ref{rem:ramp}.

\subsubsection{\bf False Alarm Performance}
The plug-in threshold of Corollary~\ref{cor:design} meets its target in 113 of
135 configurations, including all 45 at $-5$\,dB but only 26 at 15\,dB
(Fig.~\ref{fig:tail}). At low SNR, observation noise dominates the monitoring
statistic. At high SNR, variations in the WPFM estimation error account for 92 to
98\% of the window variance. The resulting statistics are right-skewed and heavy
tailed, which violates A4. Hence, calibration based on the variance
underestimates the false alarm risk, with $\ARL/\ARL^\star$ as low as 0.20.
Limited calibration data also contribute, mainly in seed~2, where the mean is
underestimated beyond $\varepsilon_{\rm cal}$. The sufficient condition of
Theorem~\ref{thm:detection}(a) fails in 60 configurations, including all 22
false alarm shortfalls (Table~S6).

\subsubsection{\bf Tail-Aware Calibration}
To address errors with heavy tails, we estimate
the adjustment coefficient from the empirical
moment generating function rather than the
variance proxy~\cite{pitts1996confidence}.
For $Z_c=S_c-\mu_0+\varepsilon_{\rm cal}-\delta/2$,
let $\hat\omega$ be the largest admissible
positive value for which the
$(1-\alpha)$ bootstrap upper confidence bound
on $\ln(C^{-1}\sum_c e^{\omega Z_c})$
is nonpositive. We set:
\begin{equation}
h=\frac{\ln\ARL^\star}{\omega},
\qquad
\omega=\min\{\omega^\star,\hat\omega\}.
\label{eq:tail}
\end{equation}
Here, $\omega^\star$ is the plug-in coefficient.
The calibration never lowers the plug-in threshold but gives no distribution-free guarantee.

With $\alpha=0.05$, 600 bootstrap resamples, and 100 calibration windows, the
tail-aware threshold meets the target in 128 of 135 configurations, including
41 of 45 at 15\,dB. The minimum $\ARL/\ARL^\star$ improves from 0.20 to 0.45.
All seven remaining failures occur in seed~2, where calibration underestimates
the mean and, at high SNR, the right tail. With 200 calibration windows (and 400
test windows), all 135 configurations meet the target, compared with 123 under
the plug-in design. Five cases meet it only narrowly.

The threshold increases by a median factor of 1.83 at 15\,dB and 1.19 at
$-5$\,dB. The detection delay increases by a median factor of 1.22 in the
reference run. Thus, tail-aware calibration improves false alarm control at the
expense of detection delay, particularly where the plug-in threshold already
meets its target.

\begin{figure}[t]\centering\input{figs/fig_v02_trace.tex}
\caption{NMSE over the switch schedule (reference run, 0.5\,dB). Main panel: trailing moving averages over 10 windows. Gray: regimes not seen in fine-tuning. Insets: values in individual windows around the change to CDL-C and the return to CDL-A. $^\dagger$Best of LMMSE and the frozen WPFM in each window (not deployable).}\label{fig:trace2}\end{figure}
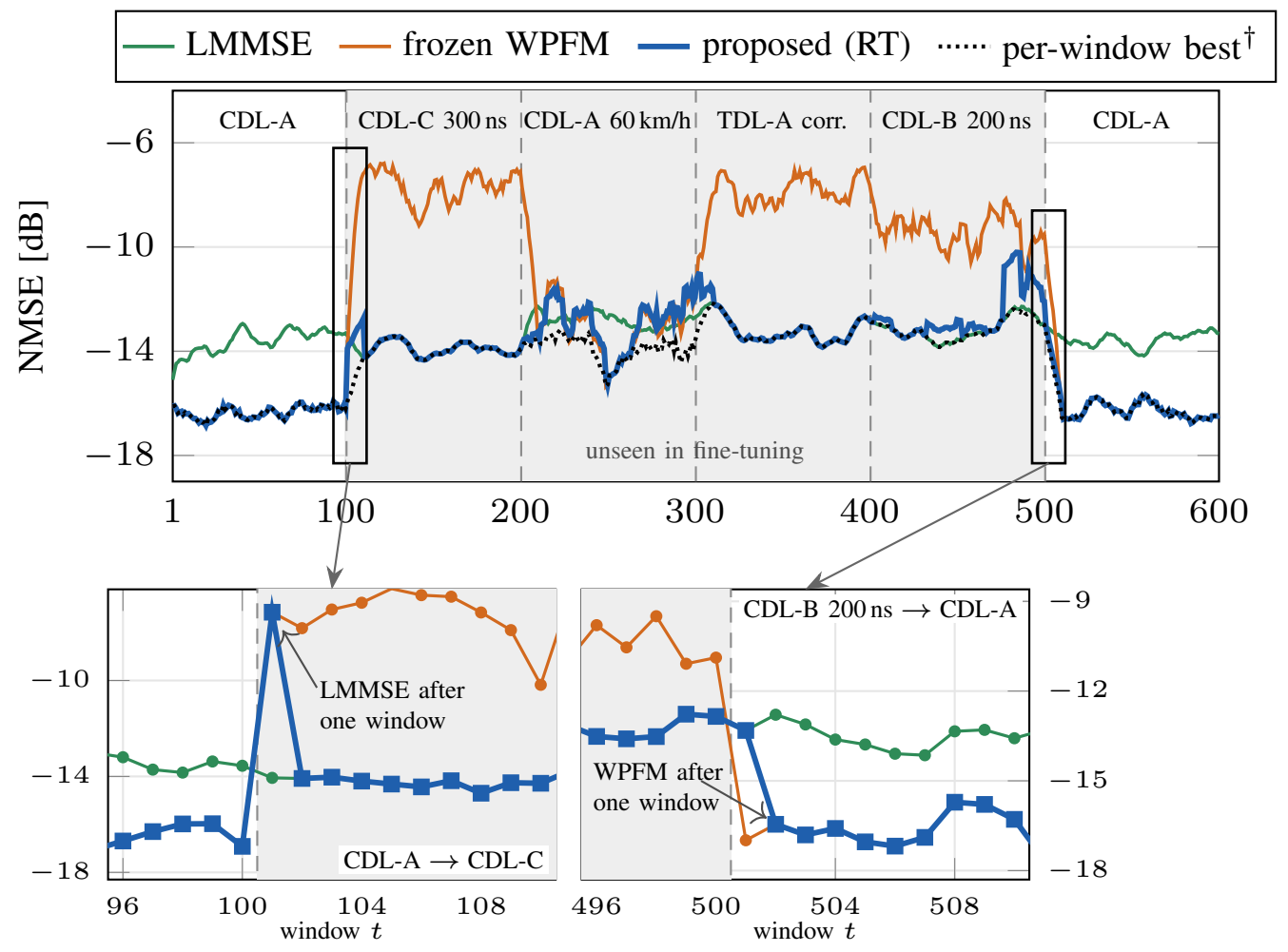

\begin{figure}[t]\centering\input{figs/fig_v02_scorecard.tex}
\caption{NMSE regret [dB] in each phase of the four end-to-end schedules, relative to the better of LMMSE and the frozen WPFM in that phase (means over five runs, negative: better than both). Last column: worst phase (mean over runs of each run's largest regret). $^\dagger$Not deployable.}\label{fig:score}\end{figure}
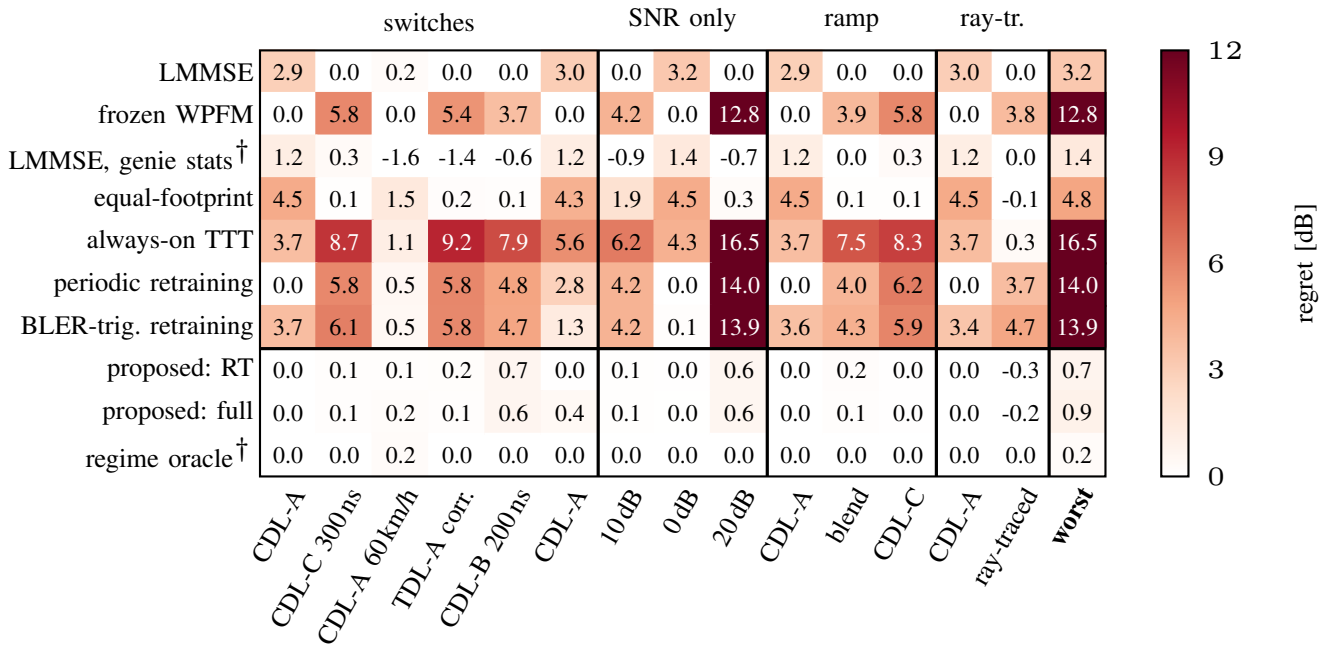

\subsection{End-to-End Estimator Selection and Adaptation}
\label{sec:res-e2e}

\subsubsection{\bf Tracking the better estimator}
Fig.~\ref{fig:trace2} shows RT selection under abrupt regime changes. On CDL-A at
0.5\,dB, WPFM outperforms LMMSE by about 3\,dB and remains selected. When the channel
switches to CDL-C (300\,ns) at window 101,
WPFM becomes 6.3\,dB worse than LMMSE and the selector switches at window 102. It returns to WPFM one window after CDL-A resumes at window 501.
Across five runs, the selector tracks the better estimator within one window of each change, except for a two-window delay in seed~3.

\subsubsection{\bf Performance relative to baselines}
Fig.~\ref{fig:score} shows the NMSE regret in each phase of the four scenarios. Fixed estimators perform inconsistently. LMMSE sacrifices up
to 3.2\,dB where the WPFM is preferable, and the frozen WPFM loses up to
12.8\,dB at high SNR. Always-on TTT and label-free retraining without fallback
lose 13.9 to 16.5\,dB in the worst phase, reflecting the risk of unvalidated adaptation. Supervised estimators also lose 4.3 to 4.5\,dB on CDL-A at low SNR.

In contrast, RT selection limits its regret in the worst phase to 0.54 to
1.07\,dB across five runs (0.63 to 1.11\,dB for the full framework). Most of the residual regret arises from the decision delay of one window and from channel variability between windows. A genie that picks the
better estimator in each window agrees with the selector in 88 to 99\% of
windows. The selector's NMSE stays within 0.03 to 0.53\,dB of the genie's. On the ray-traced trajectory, adaptive switching can even outperform a reference fixed for the whole phase, as the preferred estimator changes within it.

\subsubsection{\bf Contribution of slower adaptation loops}
The base adapter remains the best library entry across all five runs, limiting the benefit of retrieval. RT selection with retrieval and the full framework differ from RT selection alone by at most 0.10\,dB in mean
NMSE (0.36\,dB in one run). RT with TTT differs by at most 0.24\,dB, with a
0.53\,dB improvement in one run (Section~\ref{sec:res-other}). The near-RT
detector produces no alarms before the first regime change. It detects four degrading transitions within 2 to 36 windows, although
the transition to correlated TDL-A is not consistently detected. RT estimator selection provides the principal gains, whereas
retrieval and adaptation add little in the evaluated scenarios.

\begin{table}[t]\centering\caption{Ranking by MSE or by post-combining error: BLER on the four schedules (MCS~20, 10 transport blocks per window) and share of the windows of the switch schedule on the WPFM (directional run, ranges over five runs). Ranking by $S$ gives higher BLER than LMMSE on every schedule.}\label{tab:dir}\scriptsize\setlength{\tabcolsep}{3pt}
\begin{tabular}{@{}lccccc@{}}\toprule
& \multicolumn{4}{c}{BLER} & On WPFM\\\cmidrule(lr){2-5}
Deployed & Switches & SNR only & Ramp & Ray-traced & [\%]\\\midrule
\input{figs/data/v02_dir_table.tex} 
\bottomrule
\multicolumn{6}{@{}p{\columnwidth}@{}}{\scriptsize $^\dagger$Deploys, in each window, the candidate with the smaller true post-combining error (not deployable).}
\end{tabular}\end{table}

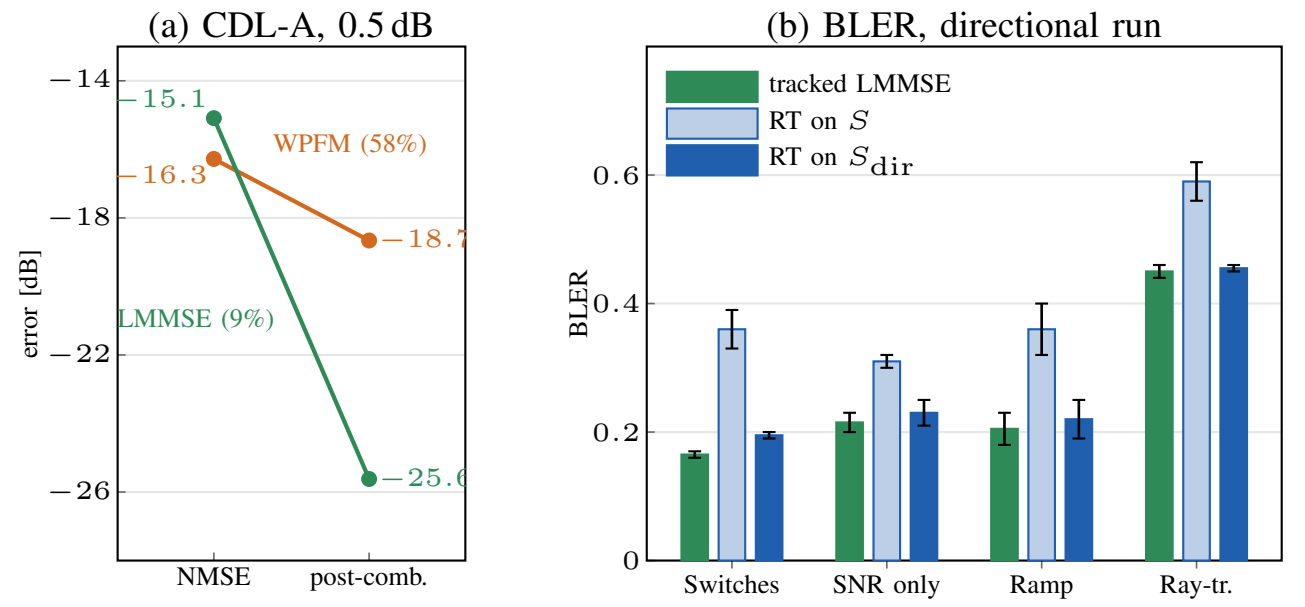
\begin{figure}[t]\centering\input{figs/fig_v02_msebler.tex}
\caption{MSE versus post-combining error. (a) CDL-A, 0.5\,dB (brackets: share of error energy along the true channel). (b) BLER by deployed estimator (bar: middle, whiskers: ends of the range over five runs in Table~\ref{tab:dir}).}\label{fig:msebler}\end{figure}

\begin{figure}[t]\centering\input{figs/fig_v02_link.tex}
\caption{Link map on CDL-A (reference run, one PUSCH transport block per slot and MCS, 10 windows per point, scored from the second). (a) BLER with the WPFM's estimate minus BLER with LMMSE's (blue: WPFM better) and the highest MCS with at most 10\% BLER for the WPFM (orange), LMMSE (green), and the true channel (dashed). (b) Throughput with the best MCS at each SNR, also for the RT selectors (inset: cell edge).}\label{fig:link}\end{figure}
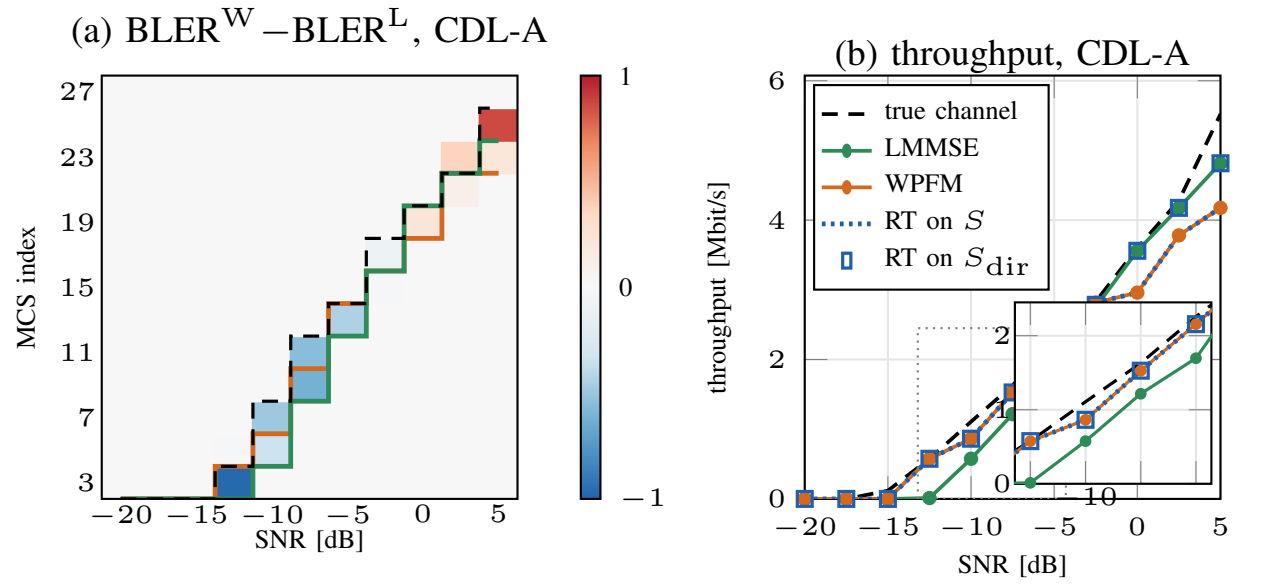

\subsection{Receiver-Aware Monitoring and Link-Level Performance}
\label{sec:res-link}

\subsubsection{\bf MSE versus decoding performance}
Lower channel-estimation MSE does not necessarily improve decoding. On the
switch schedule, selection using $S$ achieves a lower NMSE ($-13.9$ to
$-13.8$\,dB) than LMMSE ($-13.4$ to $-13.3$\,dB) but a higher BLER (0.33 to 0.39
versus 0.16 to 0.17 in Table~\ref{tab:dir}). This discrepancy arises because the
receiver treats the estimated channel as exact, which makes errors along the combining direction especially harmful. On CDL-A at 0.5\,dB, 58\% of the WPFM error energy lies along the true channel direction, versus 9\% for LMMSE. The WPFM thus has a higher post-combining error despite a lower NMSE (Fig.~\ref{fig:msebler}(a)). MSE monitoring is accurate but does not fully capture receiver performance. Selection by $S$ gave higher BLER than LMMSE on every schedule (Fig.~\ref{fig:msebler}(b) and Table~\ref{tab:dir}).

\subsubsection{\bf Directional error monitoring}
To account for receiver sensitivity, we define
a label-free directional statistic using
$\mathbf u_p=\hh^{(\cV)}_p/
\|\hh^{(\cV)}_p\|$:
\begin{equation}
S_{\rm dir}=
\frac1n\sum_{p\in\cP_h}
\frac{
|\mathbf u_p^{\mathsf H}
(\hh^{(\cV)}_p-\htl_p)|^2-\hat\sigma^2
}{
\|\hh^{(\cV)}_p\|^2
}.
\label{eq:Sdir}
\end{equation}
Under independent held-out noise and conditionally unbiased estimation of the
noise variance, $S_{\rm dir}$ is conditionally unbiased for the mean normalized
directional error power. It is a receiver-aware error proxy rather than a direct predictor of BLER. Its relation to BLER depends on combining and decoding. The concentration and paired comparison results extend directly when candidates share the same direction and normalization.
When each candidate has its own direction, the paired statistic is neither exactly Gaussian nor independent of the noise variance estimate.
The extra computation is one inner product and norm per held-out RE and candidate.

\subsubsection{\bf Receiver-aware estimator selection} 
Across all schedules and runs, the mean of $S_{\rm dir}$ differs from its target
by at most 3.5\%. Its ranking agrees with the true post-combining error in 93 to
98\% of windows. On the switch schedule, directional selection deploys the WPFM
in 5.2 to 6.7\% of windows, compared with 47.0 to 49.2\% for selection by $S$.
This lowers the BLER to 0.19 to 0.20, from 0.33 to 0.39 for selection by $S$.
The BLER remains above that of LMMSE and of the receiver-aware genie (0.16 to
0.17). Directional selection does not consistently match LMMSE. On the
other schedules, it remains within 0.02 BLER of LMMSE, compared with gaps of 0.08 to 0.19 for selection by MSE (Table~\ref{tab:dir}). The tradeoff is a smaller NMSE gain, since directional selection optimizes the post-combining error rather than the total error.

\subsubsection{\bf Link-Level Throughput}
Fig.~\ref{fig:link} compares the link-level performance of the WPFM and LMMSE
across SNRs and MCS values on CDL-A. At low SNR ($-10$ to $-5$\,dB), the WPFM
achieves up to 50\% higher throughput than LMMSE and comes within 7\% of the
reference with the true channel at $-7.5$ and $-5$\,dB. At SNRs of 0\,dB and
above, however, its throughput is up to 17\% lower, even though its NMSE remains
better up to 2.5\,dB. With selection by directional error rather than total MSE, $S_{\rm dir}$ keeps throughput within 1.2\% of the better estimator across all tested CDL-A SNRs and retains the WPFM advantage at low SNR. Similar gains appear on the unseen CDL-E, where the WPFM
achieves up to 53\% higher throughput. On CDL-C (300\,ns), LMMSE performs
better. Across these three regimes, directional selection limits the throughput
loss relative to the better estimator to 11\%. The exception is the CDL-C
coverage edge ($-12.5$\,dB), where the loss reaches 39\% of 0.2\,Mbit/s. By contrast, selection by MSE loses up to 29\% (Fig.~S11).

\subsection{Adaptation, Robustness, and Computational Cost}
\label{sec:res-other}

\subsubsection{\bf Effectiveness of label-free adaptation}
Self-supervised TTT did not improve estimation accuracy in any run. After 200
steps, the NMSE increased by factors of 1.2 to 5.4, which violates condition B1.
Retaining the current adapter generally outperformed TTT. The acceptance test of Corollary~\ref{cor:ttt} and RT selection limited the damage. Of 147 attempted updates, 91 were rejected, and RT+TTT was at
most 0.4\,dB worse than RT alone across all schedules and runs.

The four tested adapters trained without labels for individual regimes also
performed worse than the base adapter on their own regimes, with 2 to 133\% higher NMSE. Nevertheless, retrieval identified the library entry with the lowest NMSE in 96 to 100\% of trials with 50 buffered slots,
consistent with Corollary~\ref{cor:buffer}. Retraining improved the NMSE by at
most 0.18\,dB. Always-on TTT improved it by 2.0 to 2.6\,dB only on the
ray-traced trajectory and increased the BLER across all schedules. No adaptation strategy consistently outperformed RT fallback. The slower loops thus mainly rejected ineffective updates.

\subsubsection{\bf Robustness and theoretical validation}
The empirical selection error satisfies
Theorem~\ref{thm:selection}(a) in 414 of 420 evaluated cells. The six exceptions
are within Monte Carlo error (Fig.~S9). Noise variance ablations (Section~S-VIII-G) show that underestimating $\sigma^2$ by 1\,dB raises the smallest detectable shift from about 1.2 to 2.0. Underestimating it by 3\,dB raises it to between 3.1 and 3.2. Sufficiently large
overestimation makes $\hat\mu_0<0$, invalidating the threshold design and preventing detection. Thus, absolute monitoring requires the bias $|b_\sigma|$ of
the noise variance estimate to remain well below $\mbh$. The false alarm target
is met with up to eight held-out tokens and all four hold-out patterns across runs.

\subsubsection{\bf Computational feasibility}
On the CPU, deployment and monitoring require 5.2 and 5.6\,ms per slot
(Table~\ref{tab:timing}), exceeding the 0.5\,ms slot duration. Feasibility in real time on the O-DU therefore rests on analytical accelerator projections (Table~\ref{tab:compute}) and needs hardware validation. The costs of CUSUM and adapter switching are negligible at near-RT timescales.

\section{Discussion and Limitations}
\label{sec:limits}

\subsubsection{ \bf Alternative monitoring signals}
Monitoring based on BLER and KPMs detects performance degradation only after it
affects decoding and may be influenced by interference, scheduling, and link
adaptation. These KPIs are also interdependent through the shared wireless
medium in ways that are hard to simulate. In our experiments, the
BLER trigger fired periodically even without channel changes. Input and activation drift detectors~\cite{ali2026cognitive} identify distribution changes
but cannot reliably distinguish harmful from benign shifts
(Section~\ref{sec:res-harm}). They may still complement direct error monitoring.

\subsubsection{ \bf Noise variance sensitivity}
Absolute monitoring depends on the accuracy of $\hat\sigma^2$
(Lemma~\ref{lem:unbiased}), and interference or leakage on guard or null REs can bias it. Calibration can absorb a constant bias, but large errors
relative to $\mbh$ compromise detection (Section~\ref{sec:res-other}). In
contrast, the common noise variance cancels in paired comparisons of $S$ (Lemma~\ref{lem:paired}), making RT selection insensitive to this bias. This cancellation generally fails when $S_{\rm dir}$ uses a direction per candidate.

\subsubsection{ \bf Practical limitations}
The analysis assumes independent pilot noise (A1 and A2). Colored interference requires covariance estimation and whitening. Multi-user DMRS
multiplexing, including port leakage under high Doppler, remains
untested. Downlink and FDD operation require monitoring at the UE. Real-time execution on the O-DU relies on analytical projections, and the required dApp framework is not yet standardized.

All evaluations use simulation, including the ray-traced scenario. The
WPFM was trained on CPUs and is small. Several experiments across conditions use a single run, and the five runs share some channel
realizations. Stronger models may change the observed estimator crossovers.
Detection under persistent channel trajectories also remains untested. Tail-aware
calibration needs enough windows to characterize the error distribution. Seven
configurations missed their targets with 100 calibration windows but none with
200. Finally, re-baselining after an alarm uses a buffer of three windows and
keeps the initial variance estimate, causing extra alarms. Re-estimating
both statistics over a full calibration window is future work.

\section{Conclusion}

We proposed a label-free monitoring framework for wireless
physical-layer foundation models  using withheld pilots
to estimate channel-estimation error without ground-truth CSI.
We established statistical guarantees for error estimation,
drift detection, and estimator selection under explicit
assumptions, and integrated monitoring, fallback, and
adaptation across O-RAN timescales. The guarantees concern
held-out error, while deployed estimation error and BLER
are evaluated empirically.
Simulations showed that estimator performance varies with
SNR and propagation conditions. The proposed monitor
identified the lower-error estimator in 124 of 130 conditions,
while real-time fallback achieved at most 1.1\,dB phase-level
NMSE regret. Tail-aware calibration met the false-alarm target
in all 135 cases with 200 calibration windows. Although
MSE-based selection could increase BLER relative to LMMSE,
directional monitoring substantially reduced this gap.
Online adaptation provided no consistent additional benefit.

Overall, the framework enables performance-aware monitoring
and estimator fallback without ground-truth CSI. Future work
will address real-time O-DU validation, persistent-channel
operation, and receiver-aware training.

\appendices
\section{Proofs for Section~\ref{sec:stat}}\label{app:stat}
\emph{Lemma~\ref{lem:unbiased}.} By A2, $\hh^{(\cV)}_p$ is $\cF$-measurable. Given the pattern, the held-out noise is independent of $\mathbf H$, the other noise, and the model (A1, A2), i.e., $\ntl_p\perp\cF$. Since $\hh^{(\cV)}_p-\htl_p=\mathbf e_p-\ntl_p$,
\begin{equation}
\|\mathbf e_p-\ntl_p\|^2=\|\mathbf e_p\|^2-2\Re(\mathbf e_p^{\mathsf H}\ntl_p)+\|\ntl_p\|^2.\label{eq:expand}
\end{equation}
Given $\cF$ the cross term has mean zero and $\E[\|\ntl_p\|^2\mid\cF]=M\sigma^2$. Averaging gives $\E[S+\hat\sigma^2\mid\cF]=\mh+\sigma^2$, the first claim, which implies the conditional equivalence. The tower property with $\E\mh=\mbh$ and $\E\hat\sigma^2=\sigma^2-b_\sigma$ gives $\E S=\mbh+b_\sigma$ and the unconditional one. $\square$

\emph{Lemma~\ref{lem:conc}.} (a) From~\eqref{eq:expand}, $S^\circ-\mh=G+C$ with $G=-\frac2N\sum_p\Re(\mathbf e_p^{\mathsf H}\ntl_p)$ and $C=\frac1N\sum_p(\|\ntl_p\|^2-M\sigma^2)$, here with $N=nM$. Given $\cF$, $\Re(\mathbf e_p^{\mathsf H}\ntl_p)\sim\mathcal N(0,\sigma^2\|\mathbf e_p\|^2/2)$ independently over $p$, hence $G\sim\mathcal N(0,2\sigma^2\mh/N)$. Also, $\sum_p\|\ntl_p\|^2=\frac{\sigma^2}2\chi^2_{2N}$ gives $\Var C=\sigma^4/N$, and $G$, $C$ are uncorrelated because odd moments of circular Gaussians vanish. The Gaussian tail with $s=2\sigma\sqrt{\mh x/N}$ has probability at most $2e^{-x}$. The Laurent--Massart bounds~\cite[Lemma~1]{laurent2000adaptive}, $\Pr(X-D\ge2\sqrt{Dx}+2x)\le e^{-x}$ and $\Pr(D-X\ge2\sqrt{Dx})\le e^{-x}$ for $X\sim\chi^2_D$, applied with $D=2N$ and scaled by $\sigma^2/(2N)$, bound $|C|$ by $\sigma^2(\sqrt{2x/N}+x/N)$ outside $2e^{-x}$. A union bound gives $1-4e^{-x}$, and the guard estimator $\hat\sigma^2-\sigma^2=\frac{\sigma^2}{2k_g}(\chi^2_{2k_g}-2k_g)$ adds two events. Over a window, the held-out noise of the $W$ slots is independent of $\cF$ and across slots (A1, A2), and the same decomposition holds with $N=nMW$ and $\hat m_{h,t}$.
(b) The law of total variance gives $\Var S^\circ_t=\E\Var(S^\circ_t\mid\cF)+\Var\E[S^\circ_t\mid\cF]=(\sigma^4+2\sigma^2\mbh)/(nMW)+\Var\hat m_{h,t}$, and an independent $\hat\sigma^2_t$ adds its variance. For a stationary slot sequence with variance $v_m$ and autocorrelation $r_m$, $\Var(\frac1W\sum_w\hat m_{h,w})=\frac{v_m}W(1+2\sum_{\tau=1}^{W-1}(1-\frac\tau W)r_m(\tau))=:v_m/W_{\rm eff}$. $\square$

\emph{Corollary~\ref{cor:n}.} With $\hat m_h=\eta\sigma^2$ and $u=\sqrt{x/N}$, the bound of Lemma~\ref{lem:conc}(a) over $\sigma^2$ is $(2\sqrt\eta+\sqrt2)u+u^2$. It is at most $\epsilon\eta$ if $(2\sqrt\eta+\sqrt2)u\le\epsilon\eta/2$, i.e., $N\ge\frac{4x}{\epsilon^2}(\frac2{\sqrt\eta}+\frac{\sqrt2}\eta)^2$, and $u^2\le\epsilon\eta/2$, i.e., $N\ge2x/(\epsilon\eta)$. The first implies the second when $\epsilon\eta\le4$. $n_{\rm sat}$ equates the two variance terms of Lemma~\ref{lem:conc}(b). $\square$

\emph{Lemma~\ref{lem:paired}.} The $\|\ntl_p\|^2$ terms cancel:
$\|\mathbf e^A_p-\ntl_p\|^2-\|\mathbf e^B_p-\ntl_p\|^2=\|\mathbf e^A_p\|^2-\|\mathbf e^B_p\|^2-2\Re((\mathbf e^A_p-\mathbf e^B_p)^{\mathsf H}\ntl_p)$,
which is Gaussian given $\cF$ with variance $2\sigma^2\|\mathbf e^A_p-\mathbf e^B_p\|^2$, and $\mathbf e^A_p-\mathbf e^B_p=\hh^A_p-\hh^B_p$. Both statistics subtract the same $\hat\sigma^2$. Independent hold-out sets give conditional variance $(2\sigma^4+2\sigma^2(\mh^A+\mh^B))/N$. Minkowski's inequality on the empirical $L^2$ norm over $\cP_h$ gives $\sqrt{\hat\rho_{AB}}\le\sqrt{\mh^A}+\sqrt{\mh^B}$, hence the ratio bound, which is at most $2(\eta_A+\eta_B)/(1+\eta_A+\eta_B)<2$. $\square$

\emph{Proposition~\ref{prop:gap}.} For random fields on $\cD$ let $\|\mathbf X\|_\cD=(\E\frac1{|\cD|M}\sum_{r\in\cD}\|\mathbf X_r\|^2)^{1/2}$, a norm. Then $\sqrt{\mbd}=\|\hat{\mathbf H}^{(\cP)}-\mathbf H\|_\cD$, $\sqrt{\bar m^{(\cV)}_d}=\|\hat{\mathbf H}^{(\cV)}-\mathbf H\|_\cD$, and $\sqrt{\bar g_d}=\|\hat{\mathbf H}^{(\cP)}-\hat{\mathbf H}^{(\cV)}\|_\cD$. Part (i) is the reverse triangle inequality, and $\hat g_d$ uses only model outputs. (ii) $\bar m^{(\cV)}_d=\E_{r\sim U_\cD}m_r$ and $\mbh=\E_{r\sim U_{\cP_h}}m_r$. Kantorovich--Rubinstein duality gives $|\E_\mu f-\E_\nu f|\le\mathrm{Lip}(f)W_1(\mu,\nu)$ on a finite metric space, and A5 bounds $\mathrm{Lip}(m)$. (iii) $|\sqrt a-\sqrt b|^2\le|a-b|$ for $a,b\ge0$ and the triangle inequality. (iv) By (iii), $\sqrt{\mbd(r')}\ge\sqrt{\mbh(r')}-\epsilon_{\rm gap}(r')>\sqrt{\mbh(r)}+\epsilon_{\rm gap}(r)\ge\sqrt{\mbd(r)}$. Random patterns follow by averaging over the finite support. $\square$

\section{Proof of Theorem~\ref{thm:detection}(a)--(b)}\label{app:detect}
(a) Let $\omega=\omega^\star$. 
Then, under the condition in (a), $\E[e^{\omega Z_t}\mid\cF_{t-1}]\le1$. With $\Lambda_t=\sum_{j\le t}\exp(\omega\sum_{i=j}^tZ_i)=e^{\omega Z_t}(\Lambda_{t-1}+1)$ and $\Lambda_0=0$, $\E[\Lambda_t\mid\cF_{t-1}]\le\Lambda_{t-1}+1$. Hence, $\Lambda_t-t$ is a supermartingale. Since $G_t=\max\{0,\max_j\sum_{i=j}^tZ_i\}$, $\Lambda_{\tau_A}>e^{\omega h}$. If $\E_\infty\tau_A<\infty$, optional stopping at $\tau_A\wedge T$ and $\Lambda\ge0$ give $e^{\omega h}\Pr(\tau_A\le T)\le\E\tau_A$, and $T\to\infty$ gives $\E_\infty\tau_A\ge e^{\omega h}$. $\square$

(b) Since $G_t\ge G_{t-1}+Z_t$, after a change at $t_c$ with no earlier alarm, $G_t\ge\sum_{i=t_c}^tZ_i$. Hence $\tau_A-t_c+1\le T_h=\inf\{s:\sum_{i\le s}Z'_i>h\}$ with $Z'_i$ i.i.d., mean $\theta_1$, variance $v_1$, independent of the past (A4). An earlier alarm gives zero delay. Wald's identity gives $\theta_1\E T_h=h+\E R_h$, and Lorden's bound~\cite{lorden1970excess} gives $\E R_h\le\E[(Z'^+)^2]/\E Z'\le(v_1+\theta_1^2)/\theta_1$. A change inside a window adds at most one window. $\square$

\section{Proofs for Section~\ref{sec:select}}\label{app:select}
\emph{Theorem~\ref{thm:selection}.} (a) Given $\cF$, $S_c-S_{c^\star}=S_{cc^\star}\sim\mathcal N(\hat\Delta_c,2\sigma^2\hat\rho_{cc^\star}/N)$ by Lemma~\ref{lem:paired} and independence across slots. A candidate with $\hat\Delta_c>\Delta$ can be selected only if $S_{cc^\star}\le0$, which has probability $Q(\hat\Delta_c\sqrt{N/(2\sigma^2\hat\rho_{cc^\star})})\le\frac12e^{-N\hat\Delta_c^2/(4\sigma^2\hat\rho_{cc^\star})}$, where $\hat\rho_{cc^\star}>0$ for such $c$. A union bound over at most $|\cC|-1$ candidates completes (a). (b) Let $\bar c^\star=\arg\min_c\mbh^c$ and $D_c=\mbh^c-\mbh^{\bar c^\star}>\Delta$. The event $S_{c\bar c^\star}\le0$ needs either $\hat\Delta_c\le D_c/2$, with probability at most $e^{-\Delta^2/(8\nu_\Delta^2)}$ by sub-Gaussianity, or a Gaussian deviation above $\Delta/2$, with probability at most $\frac12e^{-N\Delta^2/(16\sigma^2\rho_{\max})}$. $\square$

\emph{Corollaries~\ref{cor:buffer}--\ref{cor:ttt}.} Set the bound of (a) to $\zeta$. If no candidate uses held-out observations of the buffer and the stopping slot $\tau$ depends only on $\hat\rho$, the estimates and $\tau$ are $\cF$-measurable. Given $\cF$, the held-out noise is still i.i.d.\ and (a) applies with $W=\tau$. A rule using held-out residuals would need a time-uniform bound (our buffers are fixed). For Corollary~\ref{cor:ttt}, $a'$ is trained on earlier windows and is $\cF$-measurable on the fresh hold-out set (A2). The mean of $S_{a'a}$ is $\ge0$ (resp.\ $\le-2\gamma$) and the Gaussian tail gives both bounds. $\square$

\bibliographystyle{IEEEtran}
\bibliography{refs}
\end{document}

%% file: figs/plotstyle.tex
\newif\ifillustrative \illustrativefalse
\newcommand{\illustrativestamp}{\ifillustrative
  \node[red, opacity=0.5, rotate=14, font=\bfseries\scriptsize, align=center] at (current bounding box.center)
  {REDUCED-SCALE CPU RUN\\provisional, not final results};\fi}
\definecolor{c1}{HTML}{1F5FAD}\definecolor{c2}{HTML}{D2691E}\definecolor{c3}{HTML}{2E8B57}
\definecolor{c4}{HTML}{8B3A62}\definecolor{c5}{HTML}{7F7F00}\definecolor{c6}{HTML}{5F9EA0}
\pgfplotsset{wpfmplot/.style={grid=major, grid style={gray!20}, tick label style={font=\scriptsize},
  label style={font=\scriptsize}, title style={font=\scriptsize, yshift=-1ex}, legend style={font=\scriptsize, cells={anchor=west}},
  every axis plot/.append style={line width=0.7pt}},
  changes/.style={extra x ticks=\regimechanges, extra x tick labels={},
  extra x tick style={grid=major, major grid style={black!55, densely dashed, line width=0.5pt}, tick style={draw=none}}}}
\newcommand{\loadchanges}[1]{\gdef\regimechanges{}\IfFileExists{figs/data/fig7_#1_changes.tex}{\input{figs/data/fig7_#1_changes.tex}}{}}

%% file: figs/fig1_arch.tex
\begin{tikzpicture}[
  font=\scriptsize, >=Stealth, node distance=2mm,
  box/.style={draw, rounded corners=1.5pt, align=center, inner sep=2pt, minimum height=5mm},
  loop/.style={draw, dashed, rounded corners=3pt, inner sep=3pt},
  lab/.style={font=\scriptsize\bfseries, align=left, inner sep=1pt}]
\node[box, fill=blue!6, minimum width=2.1cm] (wpfm) {WPFM $f_{\theta,\Delta_a}$\\(all pilots $\mathcal P$)};
\node[box, fill=blue!6, right=3mm of wpfm, minimum width=2.1cm] (hold) {held-out pass\\(visible $\mathcal V$): $S$, $S_{AB}$, $\hat\rho$, $\hat g_d$};
\node[box, fill=gray!8, below=2mm of wpfm, minimum width=2.1cm] (lmmse) {LMMSE (always on)};
\node[box, fill=blue!6, right=3mm of lmmse, minimum width=2.1cm] (sel) {selector (Thm.~\ref{thm:selection})};
\node[box, left=4mm of wpfm, yshift=-3.5mm, minimum width=1.1cm] (ls) {LS pilots\\$\tilde{\mathbf h}_p$, $\hat\sigma^2$};
\node[box, right=4mm of sel, minimum width=1.0cm] (dec) {equalizer,\\decoder};
\draw[->] (ls.east) -- ++(1.5mm,0) |- (wpfm.west);
\draw[->] (ls.east) -- ++(1.5mm,0) |- (lmmse.west);
\draw[->] (wpfm) -- (hold);
\draw[->] (hold) -- (sel);
\draw[->] (lmmse) -- (sel);
\draw[->] (sel) -- (dec);
\node[lab, below=1.5mm of lmmse.south west, anchor=north west] (rtlab) {RT loop: dApp in the O-DU (every $k$-th slot, $\le 1$\,ms)};
\node[loop, fit=(ls)(wpfm)(hold)(lmmse)(sel)(dec)(rtlab)] (rt) {};
\node[box, fill=orange!8, above=15mm of hold, minimum width=2.1cm] (cusum) {CUSUM on $S_t$ (Thm.~\ref{thm:detection})};
\node[box, fill=orange!8, left=3mm of cusum, minimum width=2.1cm] (retr) {retrieval (Cor.~\ref{cor:buffer}),\\TTT + rollback (Cor.~\ref{cor:ttt})};
\node[lab, right=2mm of cusum] (nrtlab) {Near-RT loop:\\xApp (10\,ms--1\,s)};
\node[loop, fit=(cusum)(retr)(nrtlab)] (nrt) {};
\node[box, fill=green!7, above=12mm of retr, minimum width=2.1cm] (lib) {adapter library $\mathcal A$,\\label-free training};
\node[box, fill=green!7, right=3mm of lib, minimum width=2.1cm] (pol) {regime declaration,\\policy $(h,\delta,W,n)$ (Prop.~\ref{prop:action})};
\node[lab, right=2mm of pol] (nonrtlab) {Non-RT loop:\\rApp ($>1$\,s)};
\node[loop, fit=(lib)(pol)(nonrtlab)] (nonrt) {};
\draw[->, thick] (hold.north) -- node[right, align=left]{E3: five scalars\\$S_t$, $S_{AB,t}$, $\hat\rho_t$, $\hat g_{d,t}$, $\hat\sigma^2_t$} (cusum.south);
\draw[->, thick] (retr.south) -- node[left, align=right]{E3/E2: adapter\\$\Delta_a$, TTT start} (wpfm.north);
\draw[->, thick] (cusum.north) -- node[right]{E2/O1: persistent alarms} (pol.south);
\draw[->, thick] (lib.south) -- node[left]{A1/O1: adapters, policy} (retr.north);
\end{tikzpicture}

%% file: figs/fig2_mask.tex
\begin{tikzpicture}[x=2.6mm, y=2.0mm, font=\scriptsize]
\foreach \l in {0,...,13} {
  \foreach \k in {0,...,23} {
    \pgfmathtruncatemacro{\isd}{(\l==2 || \l==11) ? 1 : 0}
    \pgfmathtruncatemacro{\isp}{(\isd==1 && mod(\k,2)==0) ? 1 : 0}
    \pgfmathtruncatemacro{\ish}{(\isp==1 && ((\l==2 && \k>=8 && \k<=11) || (\l==11 && \k>=16 && \k<=19))) ? 1 : 0}
    \ifnum\ish=1
      \fill[red!65] (\l,\k) rectangle ++(1,1);
    \else\ifnum\isp=1
      \fill[blue!55] (\l,\k) rectangle ++(1,1);
    \else\ifnum\isd=1
      \fill[gray!15] (\l,\k) rectangle ++(1,1);
    \else
      \fill[white] (\l,\k) rectangle ++(1,1);
    \fi\fi\fi
  }
}
\draw[step=1, gray!40, very thin] (0,0) grid (14,24);
\draw (0,0) rectangle (14,24);
\foreach \k in {4,8,...,20} \draw[gray!80] (2,\k) -- ++(1,0) (11,\k) -- ++(1,0);
\node[below] at (7,0) {OFDM symbol $\ell$ (one slot)};
\node[rotate=90, above] at (0,12) {subcarrier $k$};
\begin{scope}[shift={(15.5,15)}]
  \fill[blue!55] (0,6) rectangle ++(1,1); \node[right] at (1.2,6.5) {visible pilot $p\in\mathcal V$};
  \fill[red!65]  (0,4) rectangle ++(1,1); \node[right] at (1.2,4.5) {held-out pilot $p\in\mathcal P_h$};
  \fill[gray!15] (0,2) rectangle ++(1,1); \draw (0,2) rectangle ++(1,1); \node[right] at (1.2,2.5) {DMRS symbol, empty RE};
  \draw (0,0) rectangle ++(1,1); \node[right] at (1.2,0.5) {data RE $r\in\mathcal D$};
\end{scope}
\node[align=left, anchor=north west, text width=3.7cm] at (15.5,13.5) {Monitoring pass: the WPFM sees only $\mathcal V$. Its output at $\mathcal P_h$ is compared with the LS values there: $S=\frac{1}{nM}\sum_{p\in\mathcal P_h}\|\hat{\mathbf h}^{(\mathcal V)}_p-\tilde{\mathbf h}_p\|^2-\hat\sigma^2$.};
\end{tikzpicture}

%% file: figs/fig_v02_maps.tex
\input{figs/data/v02_cmap_labels.tex}
\input{figs/cmapstyle.tex}
\begin{tikzpicture}
\begin{axis}[cmapax, divmap, name=a, title={(a) true $\bar m_d^{\rm W}/\bar m_d^{\rm L}$ [dB]},
  yticklabels/.expand once=\cmregimes, yticklabel style={font=\tiny, inner sep=1pt}]
\cmapplot{v02_cmap_true.dat}
\cmapmark{v02_cmap_true.dat}{ns}{1}
\end{axis}
\begin{axis}[cmapax, divmap, name=b, at={(a.east)}, anchor=west, xshift=0.95cm, title={(b) label-free $\sum S^{\rm W}/\sum S^{\rm L}$},
  colorbar, colorbar style={width=0.13cm, ytick={-15,-10,-5,0,5,10,15}, yticklabels={$\le\!-15$,,,0,,,$\ge\!15$}, yticklabel style={font=\tiny}, xshift=-0.15cm}]
\cmapplot{v02_cmap_hat.dat}
\addplot[only marks, mark=x, mark size=1.4pt, black, line width=0.5pt]
  table[x=x, y=y, restrict expr to domain={\thisrow{signruns}}{0:0}] {figs/data/v02_cmap_hat.dat};
\end{axis}
\begin{axis}[cmapax, regmap, name=c, at={(b.east)}, anchor=west, xshift=0.95cm, title={(c) regret, proposed [dB]}]
\cmapplot{v02_cmap_reg_ours_K1.dat}
\end{axis}
\begin{axis}[cmapax, regmap, name=d, at={(c.east)}, anchor=west, xshift=0.95cm, title={(d) regret, SNR rule [dB]},
  colorbar, colorbar style={width=0.13cm, ytick={0,2,4,6}, yticklabels={0,2,4,$\ge$6}, yticklabel style={font=\tiny}, xshift=-0.15cm}]
\cmapplot{v02_cmap_reg_snr.dat}
\end{axis}
\begin{axis}[cmapax, kapmap, name=e, at={(a.south)}, anchor=north, yshift=-0.85cm, title={(e) MSE inflation $\kappa$ of the WPFM},
  yticklabels/.expand once=\cmregimes, yticklabel style={font=\tiny, inner sep=1pt},
  colorbar, colorbar style={width=0.13cm, ytick={0,1,2,3,4}, yticklabels={1,2,4,8,16+}, yticklabel style={font=\tiny}, xshift=-0.15cm}]
\addplot[matrix plot*, mesh/cols=13, point meta={log2(max(\thisrow{c},1))}] table[x=x, y=y] {figs/data/v02_mon_kappa.dat};
\cmapmark{v02_mon_kappa.dat}{harm}{1}
\end{axis}
\begin{axis}[cmapax, alarmmap, name=f, at={(e.east)}, anchor=west, xshift=0.95cm, title={(f) alarms, held-out $S$ (ours)}]
\cmapplot{v02_mon_alarm_S.dat}
\cmapmark{v02_mon_kappa.dat}{harm}{1}
\end{axis}
\begin{axis}[cmapax, alarmmap, name=g, at={(f.east)}, anchor=west, xshift=0.95cm, title={(g) alarms, activation drift}]
\cmapplot{v02_mon_alarm_act.dat}
\cmapmark{v02_mon_kappa.dat}{harm}{1}
\end{axis}
\begin{axis}[cmapax, alarmmap, name=h, at={(g.east)}, anchor=west, xshift=0.95cm, title={(h) alarms, input drift},
  colorbar, colorbar style={width=0.13cm, ytick={0,0.5,1}, yticklabel style={font=\tiny}, xshift=-0.15cm}]
\cmapplot{v02_mon_alarm_input.dat}
\cmapmark{v02_mon_kappa.dat}{harm}{1}
\end{axis}
\illustrativestamp
\end{tikzpicture}

%% file: figs/fig_v02_tail.tex
\begin{tikzpicture}
\begin{axis}[wpfmplot, width=\columnwidth, height=4.3cm, ymode=log, ymin=0.15, ymax=14, xmin=-3, xmax=149,
  xtick={22,73,124}, xticklabels={$-5$\,dB, 5\,dB, 15\,dB}, ylabel={$\mathrm{ARL}_0/\mathrm{ARL}^\star$},
  ytick={0.2,0.5,1,2,5,10}, yticklabels={0.2,0.5,1,2,5,10}, grid=major, xmajorgrids=false,
  legend columns=2, legend style={font=\scriptsize, at={(0.5,1.02)}, anchor=south, /tikz/every even column/.append style={column sep=4pt}}]
\draw[black, densely dashed, line width=0.6pt] (axis cs:-3,1) -- (axis cs:149,1);
\addplot[only marks, mark=o, c2, mark size=1.1pt, line width=0.4pt] table[x=x, y=r_plug] {figs/data/v02_tail_strip.dat};
\addlegendentry{plug-in proxy (Cor.~\ref{cor:design})}
\addplot[only marks, mark=diamond*, c1, mark size=1.3pt] table[x=x, y=r_tail] {figs/data/v02_tail_strip.dat};
\addlegendentry{tail-aware (bootstrap MGF)}
\node[font=\tiny, anchor=south] at (axis cs:22,0.16) {45/45$\,\rightarrow\,$45/45};
\node[font=\tiny, anchor=south] at (axis cs:73,0.16) {42/45$\,\rightarrow\,$42/45};
\node[font=\tiny, anchor=south] at (axis cs:124,0.16) {26/45$\,\rightarrow\,$41/45};
\end{axis}
\illustrativestamp
\end{tikzpicture}

%% file: figs/fig_v02_trace.tex
\begin{tikzpicture}
\begin{axis}[wpfmplot, name=main, width=\columnwidth, height=4.3cm, xmin=1, xmax=600, ymin=-19, ymax=-4,
  xtick={1,100,200,300,400,500,600}, ylabel={NMSE [dB]}, ytick={-18,-14,-10,-6},
  legend columns=4, legend style={font=\scriptsize, at={(0.5,1.02)}, anchor=south,
  /tikz/every even column/.append style={column sep=3pt}, inner xsep=1pt, inner ysep=1pt},
  legend image post style={xscale=0.6}, clip=false]
\fill[gray!13] (axis cs:100.5,-19) rectangle (axis cs:500.5,-4);
\pgfplotsinvokeforeach{100.5,200.5,300.5,400.5,500.5}{\draw[black!45, densely dashed, line width=0.4pt] (axis cs:#1,-19) -- (axis cs:#1,-4);}
\node[font=\tiny, anchor=north] at (axis cs:50,-4.1) {CDL-A};
\node[font=\tiny, anchor=north] at (axis cs:150,-4.1) {CDL-C 300\,ns};
\node[font=\tiny, anchor=north] at (axis cs:250,-4.1) {CDL-A 60\,km/h};
\node[font=\tiny, anchor=north] at (axis cs:350,-4.1) {TDL-A corr.};
\node[font=\tiny, anchor=north] at (axis cs:450,-4.1) {CDL-B 200\,ns};
\node[font=\tiny, anchor=north] at (axis cs:550,-4.1) {CDL-A};
\node[font=\tiny, gray!60!black, anchor=south] at (axis cs:300,-19) {unseen in fine-tuning};
\addplot[c3] table[x=window, y=LMMSE_s] {figs/data/v02_trace_switches.dat}; \addlegendentry{LMMSE}
\addplot[c2] table[x=window, y=WPFM_s] {figs/data/v02_trace_switches.dat}; \addlegendentry{frozen WPFM}
\addplot[c1, line width=1.1pt] table[x=window, y=RT_s] {figs/data/v02_trace_switches.dat}; \addlegendentry{proposed (RT)}
\addplot[black, densely dotted, line width=0.8pt] table[x=window, y=genie_s] {figs/data/v02_trace_switches.dat}; \addlegendentry{per-window best$^\dagger$}
\draw[black, line width=0.5pt] (axis cs:93,-18.3) rectangle (axis cs:112,-6.2); \coordinate (b1) at (axis cs:102.5,-18.3);
\draw[black, line width=0.5pt] (axis cs:493,-18.3) rectangle (axis cs:512,-8.6); \coordinate (b3) at (axis cs:502.5,-18.3);
\end{axis}
\pgfplotsset{zoomax/.style={wpfmplot, width=0.53\columnwidth, height=3.6cm,
  tick label style={font=\tiny}, label style={font=\tiny}, xlabel={window $t$}, xlabel style={yshift=1.2ex},
  every axis plot/.append style={line width=0.6pt, mark size=0.9pt}, clip=true}}
\begin{axis}[zoomax, name=z1, at={(main.south west)}, anchor=north west, xshift=-0.45cm, yshift=-0.75cm,
  xmin=95.5, xmax=110.5, ymin=-18.3, ymax=-6.2, xtick={96,100,...,108}, ytick={-18,-14,-10,-6}]
\fill[gray!13] (axis cs:100.5,-19) rectangle (axis cs:111,-5);
\draw[black!45, densely dashed, line width=0.4pt] (axis cs:100.5,-19) -- (axis cs:100.5,-5);
\addplot[c3, mark=*] table[x=window, y=LMMSE] {figs/data/v02_trace_switches.dat};
\addplot[c2, mark=*] table[x=window, y=WPFM] {figs/data/v02_trace_switches.dat};
\addplot[c1, line width=1.1pt, mark=square*, mark size=1.1pt] table[x=window, y=RT] {figs/data/v02_trace_switches.dat};
\node[font=\tiny, anchor=south east, fill=white, inner sep=1pt] at (axis cs:110.3,-18.1) {CDL-A$\,\rightarrow\,$CDL-C};
\node[font=\tiny, anchor=west, align=left, inner sep=1pt] at (axis cs:102.4,-11.0) {LMMSE after\\one window};
\draw[->, black!70, line width=0.4pt] (axis cs:102.5,-10.3) -- (axis cs:101.35,-7.9);
\end{axis}
\begin{axis}[zoomax, name=z3, at={(z1.north east)}, anchor=north west, xshift=0.18cm,
  xmin=495.5, xmax=510.5, ymin=-18.3, ymax=-8.6, xtick={496,500,...,508}, ytick={-18,-15,-12,-9}, yticklabel pos=right]
\fill[gray!13] (axis cs:490,-19) rectangle (axis cs:500.5,-7);
\draw[black!45, densely dashed, line width=0.4pt] (axis cs:500.5,-19) -- (axis cs:500.5,-7);
\addplot[c3, mark=*] table[x=window, y=LMMSE] {figs/data/v02_trace_switches.dat};
\addplot[c2, mark=*] table[x=window, y=WPFM] {figs/data/v02_trace_switches.dat};
\addplot[c1, line width=1.1pt, mark=square*, mark size=1.1pt] table[x=window, y=RT] {figs/data/v02_trace_switches.dat};
\node[font=\tiny, anchor=north east, fill=white, inner sep=1pt] at (axis cs:510.3,-8.75) {CDL-B 200\,ns$\,\rightarrow\,$CDL-A};
\node[font=\tiny, anchor=north west, align=left, inner sep=1pt] at (axis cs:495.7,-14.2) {WPFM after\\one window};
\draw[->, black!70, line width=0.4pt] (axis cs:499.5,-15.0) -- (axis cs:501.7,-16.25);
\end{axis}
\draw[black!60, line width=0.4pt, -{Stealth[length=1.6mm]}] (b1) -- (z1.north);
\draw[black!60, line width=0.4pt, -{Stealth[length=1.6mm]}] (b3) -- (z3.north);
\illustrativestamp
\end{tikzpicture}

%% file: figs/fig_v02_scorecard.tex
\input{figs/data/v02_scorecard_labels.tex}
\begin{tikzpicture}
\begin{axis}[width=0.81\columnwidth, height=4.4cm, scale only axis=false, enlargelimits=false, axis on top, clip=false,
  xmin=-0.5, xmax=14.5, ymin=-0.5, ymax=9.5, y dir=reverse,
  xtick={0,...,14}, xticklabels/.expand once=\scorephases, xticklabel style={rotate=60, anchor=north east, font=\tiny, inner sep=1pt},
  ytick={0,...,9}, yticklabels/.expand once=\scoremethods, yticklabel style={font=\tiny, inner sep=1pt},
  tick style={draw=none}, colormap={reg}{rgb255=(255,255,255) rgb255=(253,219,199) rgb255=(244,165,130) rgb255=(214,96,77) rgb255=(178,24,43) rgb255=(103,0,31)},
  point meta min=0, point meta max=12, colorbar, colorbar style={width=0.18cm, ytick={0,3,6,9,12}, yticklabel style={font=\tiny},
  ylabel={regret [dB]}, ylabel style={font=\tiny, yshift=-0.6ex}},
  visualization depends on={value \thisrow{lab} \as \celllab}, visualization depends on={value \thisrow{tc} \as \celltc},
  nodes near coords={\ifnum\celltc=1\color{white}\fi\celllab},
  every node near coord/.append style={font=\fontsize{4.5}{5}\selectfont, anchor=center, black},
  extra x ticks={5.5,8.5,11.5,13.5}, extra x tick labels={}, extra x tick style={grid=major, major grid style={black, line width=0.6pt}},
  extra y ticks={6.5}, extra y tick labels={}, extra y tick style={grid=major, major grid style={black, line width=0.6pt}}]
\addplot[matrix plot*, mesh/cols=15, point meta=explicit] table[x=x, y=y, meta=c] {figs/data/v02_scorecard.dat};
\node[font=\tiny, anchor=south] at (axis cs:2.5,-0.5) {switches};
\node[font=\tiny, anchor=south] at (axis cs:7,-0.5) {SNR only};
\node[font=\tiny, anchor=south] at (axis cs:10,-0.5) {ramp};
\node[font=\tiny, anchor=south] at (axis cs:12.5,-0.5) {ray-tr.};
\end{axis}
\illustrativestamp
\end{tikzpicture}

%% file: figs/data/v02_dir_table.tex
frozen WPFM & 0.76--0.83 & 0.30--0.33 & 0.84--0.88 & 0.73--0.78 & 100\\
tracked LMMSE & 0.16--0.17 & 0.20--0.23 & 0.18--0.23 & 0.44--0.46 & 0\\
RT on $S$ & 0.33--0.39 & 0.30--0.32 & 0.32--0.40 & 0.56--0.62 & 47.0--49.2\\
RT on $S_{\rm dir}$ & 0.19--0.20 & 0.21--0.25 & 0.19--0.25 & 0.45--0.46 & 5.2--6.7\\
per-window genie$^\dagger$ & 0.16--0.17 & 0.19--0.23 & 0.18--0.23 & 0.42--0.44 & 0.5--1.5\\

%% file: figs/fig_v02_msebler.tex
\begin{tikzpicture}
\begin{axis}[wpfmplot, name=ed, width=2.3cm, height=3.4cm, scale only axis, xmin=-0.62, xmax=1.62, ymin=-28, ymax=-13,
  xtick={0,1}, xticklabels={NMSE, post-comb.}, xmajorgrids=false, ytick={-26,-22,-18,-14}, ylabel={error [dB]},
  tick label style={font=\tiny, inner sep=1pt}, label style={font=\tiny}, ylabel style={yshift=-0.8ex},
  xtick pos=left, ytick pos=left, major tick length=1.5pt, tick style={black!60, line width=0.3pt},
  title={(a) CDL-A, 0.5\,dB}, title style={font=\scriptsize, yshift=-1.0ex},
  every axis plot/.append style={line width=0.8pt, mark size=1.1pt}]
\addplot[c2, mark=*] coordinates {(0,-16.28) (1,-18.66)};
\addplot[c3, mark=*] coordinates {(0,-15.09) (1,-25.62)};
\node[font=\tiny, anchor=north east, inner sep=1pt, c2] at (axis cs:0,-16.28) {$-16.3$};
\node[font=\tiny, anchor=south east, inner sep=1pt, c3] at (axis cs:0,-15.09) {$-15.1$};
\node[font=\tiny, anchor=west, inner sep=1.5pt, c2] at (axis cs:1,-18.66) {$-18.7$};
\node[font=\tiny, anchor=west, inner sep=1.5pt, c3] at (axis cs:1,-25.62) {$-25.6$};
\node[font=\tiny, anchor=south west, inner sep=1pt, c2] at (axis cs:0.34,-16.4) {WPFM (58\%)};
\node[font=\tiny, anchor=east, inner sep=1pt, c3] at (axis cs:0.43,-21.0) {LMMSE (9\%)};
\end{axis}
\begin{axis}[wpfmplot, ybar, area legend, name=bl, at={(ed.east)}, anchor=west, xshift=1.2cm, width=4.2cm, height=3.4cm, scale only axis,
  bar width=5pt, xmin=-0.55, xmax=3.55, ymin=0, ymax=0.8, ytick={0,0.2,0.4,0.6}, xmajorgrids=false,
  xtick={0,1,2,3}, xticklabels={Switches, SNR only, Ramp, Ray-tr.}, ylabel={BLER},
  tick label style={font=\tiny, inner sep=1pt}, label style={font=\tiny}, ylabel style={yshift=-0.8ex},
  xtick pos=left, ytick pos=left, major tick length=1.5pt, tick style={black!60, line width=0.3pt},
  title={(b) BLER, directional run}, title style={font=\scriptsize, yshift=-1.0ex},
  legend style={font=\tiny, at={(0.02,0.98)}, anchor=north west, inner sep=1pt, row sep=-2pt, draw=none, fill=none},
  every axis plot/.append style={line width=0.4pt},
  error bars/y dir=both, error bars/y explicit, error bars/error bar style={line width=0.5pt, black},
  error bars/error mark options={rotate=90, mark size=1.2pt, line width=0.4pt, black}]
\addplot[draw=c3, fill=c3] coordinates {(0,0.165) +- (0,0.005) (1,0.215) +- (0,0.015) (2,0.205) +- (0,0.025) (3,0.450) +- (0,0.010)};
\addplot[draw=c1, fill=c1!30] coordinates {(0,0.360) +- (0,0.030) (1,0.310) +- (0,0.010) (2,0.360) +- (0,0.040) (3,0.590) +- (0,0.030)};
\addplot[draw=c1, fill=c1] coordinates {(0,0.195) +- (0,0.005) (1,0.230) +- (0,0.020) (2,0.220) +- (0,0.030) (3,0.455) +- (0,0.005)};
\legend{tracked LMMSE, RT on $S$, RT on $S_{\rm dir}$}
\end{axis}
\illustrativestamp
\end{tikzpicture}

%% file: figs/fig_v02_link.tex
\input{figs/linkstyle.tex}
\begin{tikzpicture}
\begin{axis}[linkheat, linkbar, name=lk, title={(a) BLER$^{\rm W}-$BLER$^{\rm L}$, CDL-A}]
\linkheatplots{A}
\end{axis}
\begin{axis}[linktput, name=tp, at={(lk.east)}, anchor=west, xshift=1.9cm, title={(b) throughput, CDL-A},
  legend style={font=\tiny, at={(0.02,0.98)}, anchor=north west, inner sep=1pt, row sep=-2pt},
  legend image post style={xscale=0.55}]
\linktputplots{A}
\legend{true channel, LMMSE, WPFM, RT on $S$, RT on $S_{\rm dir}$}
\draw[gray, densely dotted, line width=0.4pt] (axis cs:-13.2,0) rectangle (axis cs:-4.3,2.45);
\end{axis}
\begin{axis}[linktput, at={(tp.south east)}, anchor=south east, xshift=-0.06cm, yshift=0.1cm, width=1.3cm,
  height=1.2cm, xmin=-13.2, xmax=-4.3, ymin=0, ymax=2.45, xtick={-12.5,-10,-7.5,-5}, xticklabels={,$-10$,,},
  ytick={0,1,2}, xlabel={}, ylabel={}, title={}, tick label style={font=\tiny, inner sep=0.5pt},
  axis background/.style={fill=white}, every axis plot/.append style={line width=0.6pt, mark size=0.8pt}]
\linktputplots{A}
\end{axis}
\end{tikzpicture}